\documentclass[fleqn,twoside]{article} 
\usepackage{graphicx}
\def\G{GeV$^2$}
\def\Q{$Q^2$}

 \def\vs{\vspace{.1in}}
\def\ci{\cite}
\begin{document}

\begin{flushright}
ITP-SB-97-49\\
RPI-SP-97-6
\end{flushright}
\bigskip

\pagestyle{myheadings}
%\markboth{\rm G.\ STERMAN \& P.\ STOLER}{\rm FORM FACTORS}

\centerline{\Large HADRONIC FORM FACTORS} 
\medskip

\centerline{\Large AND PERTURBATIVE QCD}
\bigskip

\centerline{\large {George Sterman}}

\medskip
\centerline{Institute for Theoretical Physics, 
SUNY, Stony Brook, NY 11794-3840}

\medskip
\centerline{\large {Paul Stoler}}

\medskip
\centerline{Department of Physics,
Rensselaer Polytechnic Institute, 
Troy, NY 12180}

\bigskip
\noindent
{\sc key words}:\quad  Electron scattering,  exclusive process, 
hadron wave function, baryon resonance, factorization,
Feynman mechanism

\bigskip
\hrule
\bigskip
\begin{abstract}
The electromagnetic form factors of hadrons 
at large momentum transfer have been the subject of 
intense theoretical and experimental scrutiny over the past two decades,
yet there is  still not  a universally-accepted framework for their description.
This review is a synopsis of their current status at large momentum transfer.
The basic theoretical 
approaches to form factors at large momentum transfer are developed,
emphasizing the valence quark and Feynman (soft) pictures.
The discussion includes
the relation of these descriptions to
the parton model, as well as 
the roles of factorization, evolution, Sudakov resummation and QCD sum rules.  
This is followed by a  discussion of the experimental status 
of pion and  nucleon elastic form factors and 
resonance production amplitudes in the light of 
recent data, highlighting the successes and shortcomings of various theoretical
proposals.
\end{abstract}

\bigskip
\hrule
\bigskip
\pagebreak
\tableofcontents

\bigskip

\section{Introduction}

\subsection{Hadronic Form Factors}
\label{HFoFa}

Exclusive electromagnetic form factors are 
a source of information about the internal structure
of hadrons.  The coupling of an elementary particle
to the photon is determined by only a few dimensionless
parameters, for example its total charge and magnetic 
moment.
For a composite particle, however, these constant coefficients
are replaced by momentum-dependent 
functions, the form factors, which
reflect the distribution of charge and current, and hence the internal
structure of the particle.  A familiar
analysis in nonrelativistic quantum mechanics relates
the electromagnetic form factor directly to the Fourier transform of the
the charge density.  Relativistic behavior also
depends very much on the nature of the hadronic state.

High momentum transfer suggests high resolution, so hard
elastic scattering is a natural way to study
the detailed internal structure of hadrons.   
Experiments in elastic electron-proton scattering 
showed long ago the famous dipole behavior of the
nucleon electromagnetic form factors
in terms of momentum transfer $Q$, $1/(1+Q^2/M^2)^2$,
with $M\sim 0.71$ \G\ \ci{Ho-56}.

Since then, many subsequent experiments studied this
and related reactions.  Their influence on our understanding
of the strong interactions themselves, however, has been somewhat overshadowed by
that of the high energy {\em inclusive} reactions.  The discovery of approximate
scaling in deeply inelastic scattering, and its
explanation in terms of the parton model, opened 
a more direct and efficient avenue to study the quarks themselves,
since inclusive rates decay much more slowly with momentum transfer.
Nevertheless, form factors at large momentum transfer remain
 an important window to quark binding in hadrons.

In this review, we will concentrate on electromagnetic 
form factors and resonance production amplitudes, at large
momentum transfer, in the light of perturbative quantum chromodynamics (QCD).
QCD itself has enjoyed so many successes, and explains so many
and varied experimental results, that it is universally
recognized as ``the" theory of the strong interactions. Yet,
the single most basic fact of the theory, the binding and confinement
of the elementary degrees of freedom, the quarks and gluons, into
hadrons, is still not described in detail.  Because of the
property of asymptotic freedom at short distances, perturbative methods must
be relevant in some degree to elastic scattering at large
momentum transfer.  Because the binding of hadrons is a
long-distance effect, nonperturbative effects must play a
crucial role as well.  The description of electromagnetic form factors requires
the consistent analysis of both length scales in a single process.  
This, and the light that will be shed on hadronic structure
by a truly successful treatment of this problem makes the study 
of form factors attractive.  We note that electromagnetic form factors
are part of the large class of exclusive hadronic amplitudes,
which also describe, for example, both proton-proton elastic scattering and 
the exclusive decays of heavy mesons.  Although 
many of the methods developed below have wide applications in this larger class,
we decided to restrict our discussion to form factors,
in the hope of improving its focus.

In the remainder of this section, we discuss what we can learn from
reasoning based on the parton model.  Here, and in most of the following,
we assume {\em very} high momentum transfer, so high that parton masses
may for the most part be neglected.  
We use parton model 
insights to identify quark counting rules, and 
as an inspiration for {\em factorization} of long- and short-distance
effects in
exclusive processes in terms of wave functions.  In this
section, we give primarily intuitive arguments, and
concentrate for simplicity on the pion.  In Sec.\ \ref{FFiQCD}, we 
discuss some of the central results of the QCD treatment of 
form factors, including the evolution of wave functions, the behavior of the 
asymptotic pion form factor, and QCD sum rules for moments of wave functions.
These topics are somewhat more mathematical, but we have attempted to
motivate technical arguments with physical intuition.  We close
this section with a brief summary  of results relevant to baryons,
especially helicity conservation, the derivation
of form factors directly from QCD sum rules, and a few phenomenological models
for moderate-$Q^2$ behavior.  We go on to review the central experimental
results for pion, nucleon and resonance production form factors, 
to assess the successes and failures 
of QCD treatments of elastic scattering, and to explain the 
controversies that have enlivened this active field of inquiry.
To anticipate, we will see that the current state of the
data is not adequate to resolve the primary
theoretical controversies.

\subsection{Partons and Factorization}
\label{PaF}

{\bf 1.2.1 Partons.} The perturbative treatment of
hard exclusive processes assumes a partonic 
description of the participating hadrons.
The general discussion is closely related
to the parton model of {\it inclusive} processes \ci{Fe-72},
such as deep-inelastic scattering.   
The celebrated premise of the parton model,
justified and systematically extended in QCD,
is that {\it inclusive} processes are determined by the
distributions $f_{i/h}(x)$, which are the
probabilities for pointlike, constituent partons $i$ to carry fraction $x$ of
the momentum of hadron $h$, summed over all other
partonic degrees of freedom.  
An {\it exclusive} form factor, on the other hand, 
reflects the coherent scattering of a hadron by an electroweak
current.  Even at large momentum transfer, it may
depend on states of definite partonic content. In
fact, at high enough energies, exclusive
amplitudes are dominated by hadronic states with ``valence"
quark content, ${\bar {\rm q}}{\rm q}$ for mesons
and qqq for baryons.  This is despite the fact
that, in its own rest frame, each hadron is
a complicated, ever-shifting superposition of 
partonic states.  Let us discuss first how such a
partonic picture of hard exclusive scattering emerges.

\begin{figure}[htbp] 
\includegraphics[angle=0,width=4in]{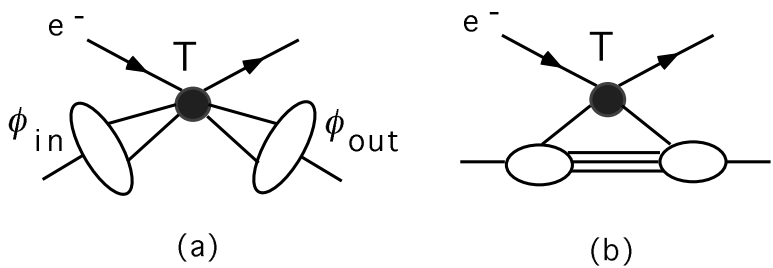}
\caption{Electron-pion elastic scattering. a) Valence PQCD picture. b)
Feynman mechanism.}
\label{epi}
\end{figure}

At sufficiently high momentum transfers
in either hadron-lepton or hadron-hadron scattering,
the {\it relative} velocities of all participating
particles are nearly lightlike.
Under this condition, the quantum
processes that bind the constituents of a
hadron are highly time-dilated in the
rest frames of the remaining particles,
both incoming and outgoing.
Correspondingly, time dilation
lengthens the lifetime of these
states, and ``freezes" the partonic
content of this hadron as ``seen" by the other particles.  
Also, as relative velocities 
approach the speed of light, the time during
which the hadrons remain in contact,
and during which momentum can be transferred, decreases.
In fact, we can always find a 
frame in which
any pair of particles 
are in contact for 
a time that decreases like $1/\gamma_{\rm rel}
=\left(1-v_{\rm rel}^2/c^2\right)^{1/2}$.
Under these conditions, we expect 
a lack of quantum interference between 
long-distance,
hadronic binding and short-distance momentum
transfer.  This {\it incoherence} between soft and hard physics
implies that we may consider each hadron
to consist of a definite partonic state during
the entire collision process.  This picture is
illustrated for electron-pion scattering in
Fig.\ \ref{epi}a, in which 
long-time dynamics, described by 
a distribution of valence quarks $\phi_{\rm in}$,
produces a ``valence"
quark-antiquark state.  
The distribution $\phi$ is often referred to as a ``wave function".
The partons of this state in turn exchange momentum
with an electron in a short-distance process
$T$.  At a later time, they reform 
a pion, through wave function $\phi_{\rm out}$.

{\bf 1.2.2 Factorization.} We summarize the above considerations 
for an arbitrary exclusive amplitude $M$ by
a schematic expression in which short-distance 
momentum transfer is factorized from the long-distance hadronic binding,
\begin{equation}
M(p_i\cdot p_j)
=
\prod_j \phi_{{\rm out},j}(n_j) \otimes
T(n_j,n_i) \otimes \prod_i \phi_{{\rm in},i}(n_i)\, .
\label{otimesfact}
\end{equation}
 Here, the labels $i$ and $j$ refer
to hadrons in the incoming and outgoing states, respectively.
$\phi(n)$ is the wave function
that describes the amplitude for a pion 
to be found in partonic state $n$, and $T(n_j,n_i)$ is
a perturbative function that describes the hard
scattering between the partons (and leptons).  The symbol
 $\otimes$ indicates a convolution, that is, a sum or integral over
the parton degrees of freedom that correspond
to states $n_i$ and $n_j$. 

A factorized expression like Eq.\ (\ref{otimesfact}) has two fundamental
properties.  First, the nonperturbative wave functions are universal
within a class of exclusive amplitudes.  This connects otherwise
disparate processes, such as the pion electromagnetic
form factor and pion-pion elastic scattering \ci{Br-89}.  
Second, the factorization of long- from short-distance
dynamics implies consistency conditions that enable us to compute
the amplitude's
dependence on the momentum transfer.  These are 
usually referred to as ``evolution" equations, examples of which we shall discuss below.
 The details of the
convolution $\otimes$, and the derivations
of evolution equations, depend on the process in question,
but one example will suffice to motivate 
Eq.\ (\ref{otimesfact}) and to illustrate the
range of possibilities, the electromagnetic form factor
of the charged pion.  We will  review the classic perturbative QCD analysis of
this form factor \cite{FJ-79,ER-80,Le-80}, and also introduce
a treatment of its ``Sudakov" effects \cite{Bo-89,Li-92}, whose importance
will become clear below.

{\bf 1.2.3 Valence PQCD and the Feynman Mechanism.}  The 
convolutions $\otimes$ in Eq.\ (\ref{otimesfact})
in principle include sums over states with arbitrary numbers 
of partons. As indicated above, however, at very
large momentum transfer, the valence state, with the
fewest partons, dominates.  We shall refer below to 
its contribution as
``valence perturbative QCD" ({\it valence PQCD}).  
This is a somewhat unconventional usage; indeed,
what we call {\it valence PQCD} is more commonly referred 
to simply as ``PQCD".  But this approximation does not exhaust the
use of perturbative methods in form factors at large
momentum transfer, and to call it simply
``PQCD" is a little misleading.  

There are, of course, many contributions 
from states with more than the valence partons.  For the most part, they
are expected to decay rapidly  with increasing momentum transfer,
relative to the valence states.  There
is an exception, however, corresponding to 
states in which one parton carries nearly
all of the hadron's momentum, while all other partons are soft.
It is plausible that such a state 
could contribute to elastic scattering, because all
of its partons except for one have long wavelengths.  They may then
overlap strongly with wave functions moving in any direction.
When the single, hard parton scatters elastically, the soft
partons from an incoming hadron may combine with the outgoing
hard parton to form an outgoing hadron.
This is illustrated for the pion electromagnetic form factor in
Fig.\ \ref{epi}b.
It is known as the
``soft" or ``Feynman" mechanism for elastic scattering.  
Nevertheless, the Feynman mechanism contains a hard scattering,
which may, in principle, be factored from the interactions
of soft partons, and treated with the methods of PQCD.  For
instance, in \cite{Du-80,Mu-81} it was analyzed for pion and nucleon
form factors.  
This PQCD investigation, unfortunately, has not yet been developed extensively
in the literature, and although it seems clear that the soft
mechanism does not contribute at asymptotically high
momentum transfer, at what scale it becomes negligible is
not well understood.  
We shall come back to the role of the soft mechanism
often below, however,
because its contribution may be studied directly in
the valence state, using Sudakov resummation, and in
QCD sum rules, and, indirectly, in models of nonperturbative hadronic
structure.

{\bf 1.2.4 The Pion Form Factor and Quark Counting.}
The electromagnetic form factor of a pion is specified by
\begin{equation}
\left( p_2+p_1\right)_\mu F_\pi(Q^2)
=
\langle \pi(p_2)|J_\mu(0)|\pi(p_1)\rangle\, ,
\label{piJme}
\end{equation}
where $J_\mu=\sum_f e_f{\bar q}_f\gamma_\mu q_f$ is the electromagnetic
current, expressed in terms of quark fields $q_f$ of flavor $f$
and electromagnetic charges $e_f$.
We neglect particle 
masses, and examine this process in a ``brick-wall" frame, in which
$p_1$ is in the plus 3 direction,
and recoils as $p_2$ in the minus 3 direction under the 
influence of the electromagnetic current $J$.  Such a
momentum configuration is most naturally described in terms of
light-cone variables, which for any vector $v^\mu$ are $v^\pm=2^{-1/2}(v^0\pm v^3)$.
In these terms we  have
\begin{equation}
p_1^+={Q/\sqrt{2}},\ p_1^-=0,\quad \quad 
p_2^-={Q/\sqrt{2}},\ p_2^+=0\, .
\label{p1p2spec}
\end{equation}
The overall momentum transfer is $(p_2-p_1)^2=-2p_1^+p_2^-=-Q^2$.

\begin{figure}[htbp]
\includegraphics[angle=0,width=4in]{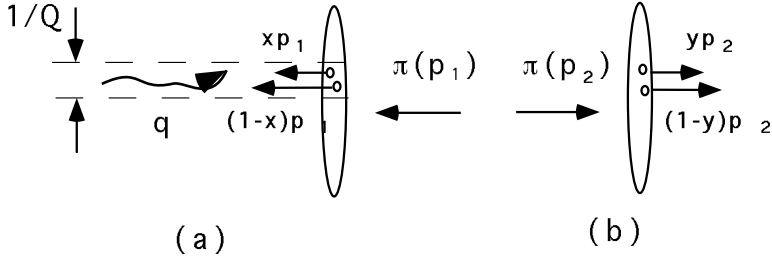}
\caption{Pion electromagnetic form factor in valence PQCD.}
\label{FormFact}
\end{figure}

The  valence PQCD portrait of this process is shown in Fig.\ \ref{FormFact}.
Fig.\ \ref{FormFact}a represents the pion in its
valence state, consisting of a quark and an antiquark.  
The variable $x$ denotes the fraction
of the pion's momentum carried by the quark and $1-x$ by
the antiquark.   
In the chosen frame, we expect the pion to be Lorentz contracted
in the direction of motion, as shown, so that the 
pair is localized in this direction.   On the other hand,
we expect the partons in any
virtual state to be more-or-less randomly distributed
in the transverse extent of the pion's wave function,
since the boost from the rest frame to the frame under
consideration leaves transverse positions unchanged.
This will have important consequences below.
Similarly, the off-shellness, and the transverse
momenta of the pair in Fig.\ \ref{FormFact} 
are boost-invariant, and we take these quantities to be fixed,
and negligible compared to both $xp_1$ and $(1-x)p_1$.
Correspondingly, the transverse components of their 
velocities vanish as $Q\rightarrow\infty$, and we neglect
them as well.  It is necessary that $1>x>0$, so that
both partons travel in the same direction as the
hadron that they represent.  
Fig.\ \ref{FormFact}a also shows
an incoming, off-shell photon, carrying momentum $q$.

Fig.\ \ref{FormFact}b
shows the state of the system after the action of the
current that absorbs the photon,
in which the pair moves in the opposite direction.
Eventually, the pair will fill out the full spatial extent
of the pion, which is again Lorentz contracted.  To form
the pion, however, their momenta must be parallel, and each
must carry a positive fraction of $p_2$, as shown.

An alternative picture relies on the ``infinite momentum frame" (IMF),
in which all participating particles move in the same direction,
with energies $E_i\gg Q$.  In this frame, all momentum transfers
are transverse.  Its main attraction lies in the conjecture that
quantization formulated in an IMF simplifies the treatment of
confinement in QCD \ci{Br-96}.

In the process depicted in Fig.\ \ref{FormFact}, the
quark undergoes a momentum transfer $xyQ^2$,  and the
antiquark $(1-x)(1-y)Q^2$, with $x$ ($y$) the fractional 
momentum of the quark in the incoming (outgoing) pion.  This must take place during
the time that the wave functions of the incoming and outgoing
pions overlap, that is, on a time scale that vanishes as $1/Q$.
The uncertainty principle requires that both members of the pair must be 
localized within $1/Q$ of each other and of the action of the current,
as indicated in Fig.\ \ref{FormFact}a.  This restriction
shows, first of all, that not all details of the valence
state wave function are relevant to exclusive scattering.
We do not need the full two-particle state;
we only need the probability for the members of the pair
to be within a transverse distance of $1/Q$ of each other.
We shall {\em assume} that this probability is
simply a function of $x$ times the geometrical factor $1/Q^2$.  This ``scaling" of
the wave function in $x$ is not exact in QCD; we
will compute corrections to it when
we discuss evolution below in Sec.\ \ref{EAB}.

Along with our assumption of
incoherence, scaling enables us to estimate the $Q$-dependence of the form
factor.   For, if long- and short-distance processes are 
incoherent, the cross section for elastic scattering of
a pion is essentially the product of the cross section for
the elastic scattering of a point-like scalar particle, times the probability
for internal processes to 
produce a virtual state in which both partons in the
valence state are within $1/Q$ of each other in transverse
distance.  
Thus, we have
\begin{equation}
\sigma_{{\rm el},\pi}\ \sim\ \sigma_{\rm el,\ point}\times F_\pi^2(Q^2)\
\sim\ \sigma_{\rm el,\ point}\times (1/Q^2)^2
\end{equation}
so that
\begin{equation}
F_\pi(Q)\sim (1/Q^2)\, .
\end{equation}
Results of this sort, based on incoherence, scaling
and geometrical estimates, are known as ``quark counting" \ci{Br-72,Ma-72}.
Quark counting rules give for an arbitrary exclusive process involving $n_h$ hadrons,
\begin{equation}
\sigma(Q^2)_{\rm had}
 = \sigma(Q^2)_{\rm point}(Q^2)^{-n_q+n_h}f\, ,
\label{qkcount}
\end{equation}
where $n_q$ is the total number of quarks and antiquarks taking
part in the process, and $f$ depends on dimensionless variables.  

>From Eq.\ (\ref{qkcount}), we see that interactions involving more than the minimum
number of partons -- say, a gluon in addition to the pair --
are suppressed by {\it a power} of $Q$, because as
$Q$ grows, the likelyhood of finding more than the minimum number 
of particles within $1/Q$ of each other falls as $Q^{-2}$ for
each additional particle.  

We note, however, that in the limits $x,y\rightarrow 0,1$, our
process describes the elastic scattering of an on-shell quark (antiquark)
with nearly all of the pion's momentum.  The remaining, soft antiquark (quark)
has long wavelength, which overlaps with both the incoming and
outgoing wave functions.  
This is the intersection of valence PQCD with the Feynman mechanism.

In the next section, we shall turn to the field-theoretic treatment
of the pion's form factor, and shall see how these features
of the parton model are realized within QCD.

\section{Form Factors in QCD}
\label{FFiQCD}

\subsection{The Factorized Pion Form Factor}
\label{FPFF}

\begin{figure}[htbp]
\includegraphics[angle=0,width=4in]{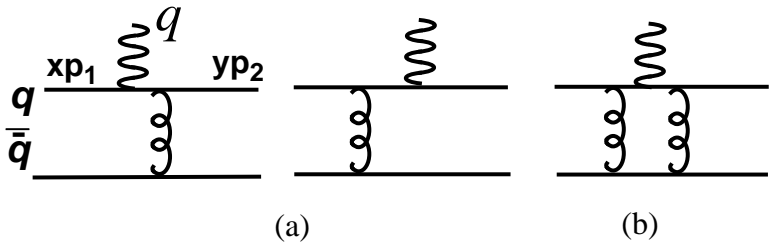}
\caption{Graphical contributions to  $T$: (a) Lowest order; (b)
Typical one-loop correction.}
\label{ffgraph}
\end{figure}

{\bf 2.1.1 Convolution in Fractional Momenta.}
We are now ready to turn to the pion form factor in valence PQCD \cite{FJ-79,ER-80,Le-80}.
The parton model discussion of the previous
section suggests that the pion
form factor can be written, following Eq. (\ref{otimesfact}),
as a sum over wave functions involving only quark momentum fraction.
We denote these as
 $x$ and $y$ 
for the incoming and outgoing pions,
respectively.  We then have the following representation 
for the form factor,
\begin{equation}
F_\pi(Q^2)=\int_0^1 dx\; dy\; \phi_\pi(y,\mu^2)\; T(y,x,Q^2,\mu^2)\; \phi_\pi(x,\mu^2)\, .
\label{pifffact}
\end{equation}
Here, $\phi_\pi (x, \mu^2)$ is the valence-state wave function
describing a quark with fraction $x$ of the pion's momentum.  $T (x, y,
Q^2, \mu^2)$ describes the hard scattering of partons.  It is a
perturbative expansion in the strong coupling at scale $\mu^2$ ($\alpha_s
(\mu^2)$), and is free of infrared divergences order-by-order in
perturbation theory.  At lowest order, it is given by the diagrams shown
in Fig.\ \ref{ffgraph}a.  
Since the incoming and outgoing pairs are
each at a tiny transverse separation, orbital angular momenta
are negligible, and partonic helicities must sum to zero.
The pairs of
incoming and outgoing external lines in the diagrams are 
thus projected onto Dirac matrices
that represent these helicity-zero pairs. \ci{Le-80}
At the same time, because masses are neglected, helicities are
conserved in perturbation theory, and hence in the hard scattering,
to all orders.
This has important consequences for
hadrons with spin.
An exercise in dimensional counting shows that $T$ has
dimensions (mass)$^{-2}$, and hence scales as $1/Q^2$.
Its explicit lowest-order form is
\begin{equation}
T_H=16\pi C_F\alpha_s(\mu^2)\left[ {2\over 3}{1\over xyQ^2}
+{1\over 3}{1\over (1-x)(1-y)Q^2}\right ]\, ,
\label{loth}
\end{equation}
where $C_F=(N^2-1)/2N=4/3$ in QCD.

In perturbation theory, 
it is possible to show that the valence PQCD result Eq.\ (\ref{pifffact})
is a theorem, which
describes the behavior of $F_\pi(Q^2)$ at large $Q^2$ 
to all orders in $\alpha_s$.  Corrections are suppressed by {\it powers}
of $Q^2$, including those due to the Feynman (soft) mechanism
described above  \cite{ER-80,Le-80,Mu-81}. Helicity conservation in $T$ is also
valid up to similar corrections.
Space does not allow a discussion here of technical aspects of this proof or of the
calculation of $T$ beyond lowest order.  
The basic technique is
already illustrated by a typical one-loop correction, Fig. \ref{ffgraph}b.
 (For the pion form factor, the full one-loop calculation
has been performed explicitly \cite{Di-84,Br-87}.  We 
keep only those contributions to $T$ from Fig.\ \ref{ffgraph}b
where all quark and gluon lines  are
{\em off-shell by at least the renormalization scale}, $\mu^2$.  
 $T$ then depends on only two momentum scales, $Q$ and $\mu$. Alternately,
we can think of $\mu^2$ as the minimum transverse momentum carried by lines in $T$.
Suppose, now, that we choose $\mu=Q$.
By the uncertainty principle, this corresponds to including in $T$ only
lines that are within $1/Q$ of each other in transverse distance, as
we anticipated in our discussion of Fig.\ \ref{FormFact} above.
By choosing $Q=\mu$ we get the extra benefit of expanding $T$ in terms of the small
parameter $\alpha_s(Q^2)$.  Thus, $\mu=Q$ will be our default choice of
scale below, although other choices may sometimes offer special advantages. 

{\bf 2.1.2 Transverse Degrees of Freedom in $F_\pi$.}
Our next exercise in factorization is to return to
Eq.\ (\ref{pifffact}), taking 
into account transverse degrees of freedom.  Remaining in
the valence picture, we recall that
the pair in the incoming and outgoing pions are not 
literally at a point, but are separated by transverse vectors
$b_i$ when they undergo the hard scattering, where $i=1\, (2)$ 
for the incoming (outgoing) pion.  Again, $Q^2=(p_2-p_1)^2$.
The wave functions in Eq.\ (\ref{otimesfact}),
which we now denote $\cal P$, are characterized by
both fractional momenta {\em and} transverse separation, and
the form factor is reexpressed as a convolution in both \ci{Li-92},
\begin{eqnarray}
F_\pi(Q^2) &=& 
\int_0^1 dxdy\; \int {d^2b_1\over (2\pi)^2} {d^2b_2\over (2\pi)^2}\; {\cal P}(y,b_2,p_2,\mu)\cr
&\ & \hbox{\hskip 1.5 true cm} \times\; 
T(y,x,p_i,b,\mu)\;
 {\cal P}(x,b_1,p_1,\mu)\, ,
\label{pilist}
\end{eqnarray}
where $T$ is a new hard-scattering function.  Beause we 
integrate over the variables $b_i$
conjugate to transverse momenta, $\mu$ does not play 
the role of a ``transverse momentum
cutoff", as in Eq.\ (\ref{pifffact}), but is simply the renormalization scale. 
On the other hand, the wave functions $\cal P$ depend
upon the momenta $p_i$ and they, along with $T$, are
not individually Lorentz invariant.  The requirement
of Lorentz invariance in the complete amplitude $F_\pi$
will lead to evolution equations below.
We emphasize that, summed to all orders, Eq.\ (\ref{pilist})
is equivalent to Eq.\ (\ref{pifffact}) at leading power in \Q .
Depending on the details of $\cal P$, however, it 
differs from Eq.\ (\ref{pifffact}) in 
nonleading powers of \Q\ in general.  

We next explore the
relation between the two factorization procedures a little further.
Intuitively, we expect that 
the wave function $\cal P$ near $b=0$, that is at
small separation for the pair, is related to the 
distribution amplitude $\phi_\pi$.  Specifically, it is not
difficult to show that \ci{Bo-89}
\begin{equation}
{\cal P}(x,b=1/\mu,p_i,\mu)\  \sim\ \phi(x,\mu^2)\, ,
\label{phitocalP}
\end{equation}
up to corrections that are suppressed by 
the strong coupling evaluated at the factorization scale $\mu$.
  The Lorentz noninvariance of 
the $\cal P$ disappears in this limit.

Let us now compare Eq.\ (\ref{pilist}) to the classic expression, Eq.\ (\ref{pifffact}).
If $Q^2$ is large enough, we expect,
according to our discussion above, that $T$ in Eq.\ (\ref{pilist}) is  
concentrated near $b\sim 1/Q$,
so that by Eq.\ (\ref{phitocalP}) $\cal P$ may be replaced by $\phi_\pi$.  
In this limit, the two expressions are  equivalent.  A closer look at 
$T$ in Eq.\ (\ref{pifffact}), however, shows that it 
actually corresponds to a localization in transverse space
only at the scale $(xyQ^2)^{-1}$.  When $x$ or $y$ vanishes,
the hard scattering ``spreads out" in transverse space, 
and violates the original assumptions of the partonic discussion of 
Sec.\ \ref{PaF}.1 above, and the reaction is defined by
the Feynman mechanism.  
Also note that if $b$ is large, the neglect of orbital contributions
to helicity is no longer justified, even if
helicity is conserved in the hard scattering \ci{JR-92,GPR-96,KW-96}.
The contribution of the ``end-point regions" $x,y\rightarrow 0,1$
(the equivalent of the Feynman mechanism for ${\rm {\bar q}q}$ states)
depends on the details of the wave functions $\phi$, but it poses
a problem, unless $Q$ is very large \ci{Is-89,Ra-84}
We shall see shortly that the use of the modified factorization in Eq.\ (\ref{pilist}) 
serves to stabilize the valence PQCD picture of scattering
at somewhat lower \Q\ than in Eq.\ (\ref{pifffact}) \ci{Li-92}.
To see how this comes about, we turn now to a discussion of evolution,
as derived from the factorization  formulas Eq.\ (\ref{pifffact})
and Eq.\ (\ref{pilist}).  

\subsection{Evolution and Asymptotic Behavior}
\label{EAB}

Eqs.\ (\ref{pifffact}) and (\ref{pilist}) for the elastic form factor
 are both convolutions of functions that depend upon arbitrary
choices:  the renormalization scale $\mu$ in the former case, and
the Lorentz frame in the latter.  In fact, a
great deal can be learned from these parameters,
through their role in the factorization formulas.
Among other things, it will allow us, in the following subsection, to give 
an explicit expression for
the asymptotic behavior of $\phi_\pi$ and
the form factor at high momentum transfer.

{\bf 2.2.1 Evolution.} Consider Eq. (\ref{pifffact}) for $F_\pi (Q^2)$.  The physical
form factor, of course, cannot
depend upon $\mu$:
\begin{equation}
\mu{d\over d\mu}F_\pi(Q^2)=0\, .
\end{equation}
Equivalently, in terms of the hard-scattering and wave functions,
\begin{eqnarray}
0 &=& \int_0^1 dx dy \bigg [ {d\phi_\pi(y)\over d\mu}T\phi_\pi(x)
 + \phi_\pi(y) {dT\over d\mu} \phi_\pi(x) \cr
&\ & \hbox{\hskip 1.5 true cm} +\ \phi_\pi(y) T {d\phi_\pi(x)\over d\mu}\;
\bigg ]\, .
\label{varyfpi}
\end{eqnarray}
This expression may be treated by separation-of-variable techniques.
$d \phi_\pi(y,\mu^2) / d \mu$, for instance, may depend upon the variables
$y$ and $\mu^2$ only, the latter only through $\alpha_s (\mu^2)$ (since
there are no other dimensionless variables available.)  In fact, its
derivative with respect to $\mu^2$ must be perturbatively calculable,
because changes in $\mu$ shift contributions from lines that are
off-shell by order $\mu^2$ between $\phi_\pi$ and $T$. (See Sec.\ \ref{FPFF}.1).) The most
general form that satisfies these requirements is itself a convolution
\ci{Le-80}:
\begin{equation}
\mu{d\phi(y,\mu^2) \over d\mu}
=
\int_0^1 dz\, V(y,z,\alpha_s(\mu^2))\phi_\pi(z,\mu^2)\, .
\label{evolve}
\end{equation}
The kernel $V$ is a distribution, rather than a simple function of
$y$ and $z$, but its integral with any smooth function is finite.  Given
the convolution form Eq.\ (\ref{pifffact}) for the form factor, the evolution equation
(\ref{evolve}) holds to all orders in $\alpha_s (\mu^2)$.  Its explicit one-loop form is
simply the coefficient of $\ln Q^2$ in the sum of one-loop
corrections to the hard scattering, such as Fig.\ \ref{ffgraph}b. The
kernel $V$ is known up to two loops \ci{Mu-95}.  We shall not exhibit its explicit
form, but only note that, with the one-loop $V$, Eq.\ (\ref{evolve}) may be
solved explicitly.
The most general solution is an expansion in
Gegenbauer polynomials $C_n{}^{3/2}$ \ci{Le-80} \ci{Ef-80},
\begin{equation}
\phi_\pi(x,\mu^2)=x(1-x)\sum_{n\ge 0}a_n C_n^{3/2}(2x-1)\big ( \ln{\mu^2\over \Lambda^2} 
\big )^{-\gamma_n/2\beta_2}\, ,
\label{evolnsoln}
\end{equation}
with $\beta_2 = (33 - 2n_f)/12$ the one-loop coefficient of the
QCD beta function, the $\gamma_n$ known anomalous dimensions and the
$a_n$ arbitrary coefficients.

Space allows us to make only a few observations on this fascinating
result:
(i) The $a_n$ are linear combinations of matrix elements,
identified in Sec.\ \ref{wavefnnp} below;
(ii) $\gamma_0 = 0$.  This is because the $n=0$ wave
function, $\phi_0
(z) = a_0 z(1-z)$ gives zero when integrated with the one-loop
kernel in Eq.\ (\ref{evolve}).  We shall refer to this
``asymptotic" form of the pion wave function many times below;
(iii) for $n>0$, all $\gamma_n > 0$, which implies that as $\mu^2 \to
\infty$, all  $x$-dependence in (\ref{evolnsoln}) that is not in the form of the
asymptotic wave function decays, albeit only logarithmically.

{\bf 2.2.2 Sudukov Resummation.} Turning now to factorization
in transverse space, we see that
the factorization Eq.\ (\ref{pilist}) suggests another
evolution equation, this time in the momentum scale $Q$,
which enters the wave functions through (non-invariant dependence on)
the momentum vectors $p_i$.
This equation
will allow us to resum perturbative logarithms of the
form $\ln(Qb)$, with $b$ the distance between the hard scatterings,
an example of ``Sudakov resummation".  
The derivation of this equation is given in a related
context in \ci{Bo-89}.
Here, we shall content
ourselves with a physical explanation and the basic results.
In brief, the effect of the resummation will be
to suppress the
nonperturbative contribution to $F_\pi$, and thus 
to extend valence PQCD to lower \Q.

For large momentum transfer, the dynamics of elastic scattering 
strongly disfavors configurations in which $b$ is large.
The physical reason for this result is that an isolated accelerated charge
must radiate, by correspondence to classical gauge theory.
 As $b$ grows, the two charges associated with
quark and antiquark become more isolated, and have correspondingly
more tendency to radiate gluons.
In elastic scattering, however,  such radiation is forbidden
by definition. 
Perturbatively, this manifests itself in 
the presence of double-logarithmic
(``Sudakov")
corrections of the form $\alpha_s^n(\mu)\ln^{2n}(bQ)$.
We therefore expect that 
the double logarithms at large $b$ will suppress configurations
for which the charges are separated far enough to 
couple strongly to radiated gluons. 
Because the effect is essentially classical, it is necessary to
sum to all orders (take the limit of large quantum numbers) to
make this suppression manifest.

In this case, an evolution equation 
is derived from the independence
of  expressions like Eq.\ (\ref{pilist}) of the choice of inertial frame.  
An infinitesimal Lorentz transformation changes the arguments
of the $\cal P$'s and of  $T$, but otherwise leaves the
amplitude invariant.
A full derivation 
(see \ci{Bo-89}; the reasoning there is an application
of a method first developed in Ref.\ 
\ci{Co-81})
requires more analysis of $b$ and $Q$ dependence
in $T$ than we have room for here.
The result is the following evolution equation, which takes the place of Eq.\ (\ref{evolve}).
Taking $p^+=Q$ in the center-of-mass frame, we have,
\begin{equation}
Q{\partial \over \partial Q}{\cal P}(x,b,p,\mu)
=\left [  K(b\mu)+G(x,Q/\mu) \right ] {\cal P}(x,b,p,\mu)\, ,
\end{equation}
in which the functions $K$ and $G$ may be computed in perturbation theory.
$K$ depends only on the ``infrared" variable $b$, and $G$ on the ``ultraviolet"
variable $Q$.

The details of the solution to this equation is straightforward,
and may be found in \ci{Bo-89}. The result is striking:
\begin{equation}
{\cal P}(x,b;p,\mu)
=
{\rm e}^{-S(x,b,Q,\mu)}\bigg ( \phi_\pi(x,1/b^2) + {\cal O}\left (\alpha_s^2(1/b)\right ) \bigg )\, ,
\label{resumP}
\end{equation}
where $\phi_\pi$ is the usual light-cone wave function for the pion,
now evaluated at $\mu=1/b$.
The Sudakov exponent $S$ strongly suppresses the wave function at large $b$,
through the summation of double logarithms of $bQ$ per loop,
\begin{equation}
S
=
C_F \int_{1/b}^{x Q} {d\mu'\over \mu'}{\alpha_s(\mu')\over\pi}\ln \left({x Q\over \mu}\right)
+\ x \leftrightarrow 1-x + \dots\, ,
\label{sudexp}
\end{equation}
where we have suppressed terms with fewer logarithms per loop.  Note in particular that
within the integral the perturbative coupling runs with the variable $\mu'$, so that
the Sudakov exponent $S$ diverges at $b=1/\Lambda_{\rm QCD}$.  
When $Q\gg \Lambda_{\rm QCD}$, the exponent
is large, and the suppression great, whenever $b\gg 1/Q$, even for $b\ll 1/\Lambda_{\rm QCD}$. 
The quark-antiquark state with opposite helicties again dominates in this limit.
The suppression of large-$b$ configurations has many applications
to hadron-hadron reactions, and helps justify the concept of
``transparency" in hadron-nucleus scattering \ci{Ja-96}.

{\bf 2.2.3 The Asymptotic Form Factor.} 
We are now ready to discuss one of the central results of the perturbative
treatment, the asymptotic behavior of the pion electromagnetic form factor.  
We begin by recalling that the natural choice of scale in 
the factorized expression Eq.\ (\ref{pifffact})
is $Q=\mu$ (See Sec.\ \ref{FPFF}.1).  For $Q$ large enough, then, the wave function will be dominated
by the $a_0$ term in its expansion (\ref{evolnsoln}), {\it and} $T$ will be
well-approximated by its lowest-order contribution, Eq.\ (\ref{loth}).  The
$x$ and $y$ integrals in (\ref{pifffact}) are then simple, and the
only remaining uncertainty is in a factor of $a_0^2$.

To fix $a_0$, we observe that the decay of the charged pion though the weak
interactions may be treated by the same method of factorizing hard and
soft degrees of freedom.  In this case, the hard interaction is at a scale
of the order of the W-mass, and the wave function of Eq.\ (\ref{evolnsoln}) is again
dominated entirely by its $a_0$ coefficient.
Then, defining the pion decay constant $f_\pi$ by (with $f_\pi\sim 93\; {\rm MeV}$)
\begin{equation}
\langle 0|{\bar d}(0)\gamma^\mu(1-\gamma_5)u(0)|\pi(p)\rangle = -\sqrt{2}p^\mu f_\pi\, ,
\end{equation}
we may identify $a_0 = \sqrt{3}f_\pi$ 
in Eq.\ (\ref{evolnsoln}). (More generally, we have $3 f_\pi
/\sqrt{N}$ with $N$ the number of colors).  This result, along with
properties of the anomalous dimensions ($\gamma_n>0$ for $n>0$), allows us to identify
the large
$\mu^2$ (or $Q^2$) behavior of the pion's quark wave function:
\begin{equation}
\phi_\pi(x,\mu^2) \rightarrow \sqrt{3}f_\pi x(1-x)\, .
\label{asywf}
\end{equation}
This is generally referred to as the {\it asymptotic} wave function of the pion.
We emphasize that it is model-independent.

Substituting Eq.\ (\ref{asywf}) into Eq.\ (\ref{pifffact}), 
and using the lowest order hard-scattering function $T$ (Eq.\ (\ref{loth})) with
$\mu=Q$, we find an elegant
expression for the pion form factor at high energy,
which is valid up to corrections in $\alpha_s(Q)\sim 1/\ln(Q)$ \cite{ER-80,Le-80},
\begin{equation}
F_\pi(Q^2)
=
{12f_\pi^2\pi C_F\alpha_s(Q^2) \over Q^2}\, .
\label{fpiasym}
\end{equation} 

{\bf 2.2.4 Sudakov Resummation for $F_\pi$.}
With an eye to contributions for
which the pair is widely  separated, we may also use  the Sudakov-resummed 
transverse wave function (\ref{resumP}) in Eq.\ (\ref{pilist}), to get
\begin{eqnarray}
F_\pi(Q^2) &=& 
\int_0^1 dxdy\; \int {d^2b\over (2\pi)^2}\; \phi_\pi(y,1/b^2)\; {\rm e}^{-S(y,b,Q,\mu)}\cr
&\ & \ \times\; 
T(y,x,Q,b,\mu)\;
{\rm e}^{-S(x,b,Q,\mu)}\; \phi_\pi(x,1/b^2)\, ,
\label{listresum}
\end{eqnarray}
where we have simplified to a single transverse separation
\ci{Li-92,Hu-96}.
The Sudakov exponential in this expressions
factor suppresses contributions from $b\gg 1/Q$.  The
natural scale of the coupling in $T$ is $\mu\sim 1/b$, even
in the end-point region.  Perturbation theory thus remains 
self-consistent, by the dynamical suppression of the 
overlap region of valence PQCD and the soft mechanism.  For moderate
\Q, however, $F_\pi$ still receives substantial
contributions from relatively large $b<1/\Lambda_{\rm QCD}$.
In this region, (\ref{listresum}) should be thought of as
a valence PQCD {\em model} for $F_\pi$.  Form factors
computed according to each of these procedures will be
confronted with the data in Sec.\ \ref{ESHFF} below.  

\subsection{Wave Functions and Nonperturbative Analysis}
\label{wavefnnp}

In the factorized picture of elastic scattering we
treat hadrons as superpositions of 
states, each with definite numbers and positions
(or momenta) of partons.
Also, as we have seen, it is the
states with the fewest partons - the ``valence
states" that dominate exclusive
processes at sufficiently high \Q\ .  
Relativistic valence
wave functions for valence states may be
identified with matrix elements that connect
single-particle states of definite hadron momentum
$|h(p)\rangle$, with the hadronic vacuum
$|0\rangle$ by the action of fields that absorb
 the relevant valence quanta.  
The analysis of
these matrix elements can lead to valuable nonperturbative
information, which supplements the purely perturbative
results outlined above.  

{\bf 2.3.1 Matrix Elements.} 
The light-cone wave function 
in position space for the valence state of a $\pi^+$ 
may be defined in terms of the matrix element of
an up quark field $u$ with a conjugate down quark field,
\begin{equation}
\Psi(z\cdot p,z^2)
=
\langle 0|
\bar{d}(0)\; \gamma^+\gamma_5 u(z)
|\pi^+(p)\rangle\, ,
\label{piPsi}
\end{equation}
where $p$ is taken in the plus direction. 
In the following, we shall
generally 
neglect hadronic masses.
We recall that
$\gamma^\pm={1/\sqrt{2}}\left (\gamma^0\pm
\gamma^3\right)$.
So that $\Psi(z\cdot p,z^2)$ may have a
natural interpretation in terms of 
independent measurements of the up and antidown
quark fields, we choose the separation between
the two fields to be spacelike,
$z^2<0$.  The Dirac structure $\gamma^+\gamma_5$ 
projects out precisely the zero-helicity combinations
of the quark and antiquark fields.

As defined, 
the wave function $\Psi$ of Eq.\ (\ref{piPsi})
is gauge-dependent.  A common choice of
gauge for the gluon field $A$ is $A^+=0$ for a
pion moving in the plus direction.   
Alternately, we may connect the fields ${\bar d}(0)$ and $u(z)$
by a path-ordered exponential in the direction $z^\mu$,
$P\exp\left[\int_o^1 dt z^\mu A_\mu(zt)\right]$, with 
$A_\mu$ expressed as a matrix in the quark
representation.

The momenta of the partons in the
valence state may be fixed by taking Fourier
transforms.  
For $\phi_\pi(x)$, we fix the fractional momentum of the quark to be $xp$ in the pion's
direction of motion, and integrate freely over all of its other
components (and hence those of the antiquark) by setting $z^+=z_T = 0$, and taking the transform 
of $\Psi$ with respect
to $z^-$,
\begin{equation}
\phi_\pi(x,\mu^2)=\int_{-\infty}^\infty{dz^-\over2\pi}\; 
{\rm e}^{iz^-xp^+}\Psi(z^-p^+,0)_{A^+=0}\, .
\label{phidef}
\end{equation}
Here $\phi_\pi(x)$ depends explicitly on the renormalization scale
$\mu^2$, because the limit $z^+,z_T \to 0$, which takes $z^\mu$ to the
light cone, $z^2 = 0$, is singular.  Defined in this
fashion, $\phi_\pi(x)$ is referred to as a {\em light cone wave function}.

Readers familiar with the QCD analysis of deeply inelastic scattering (DIS)
will recognize a similarity between the valence quark wave function
given by Eqs. (\ref{piPsi}) and (\ref{phidef}) and the inclusive parton distribution 
density in a hadron. 
Note, however, that while a parton distribution in DIS is a probability,
$\phi_\pi$ is an amplitude.
Thus, although we know the behavior of
our light-cone wave functions at very large $\mu$, 
they might evolve slowly to this form, and we would like further
information on their properties for intermediate values of $\mu$.
A direct approach is to compute the relevant matrix elements using
the methods of lattice QCD.  Moments
of proton wave functions have been computed in
this fashion \ci{Ma-89,Ne-95}, and more work may be antitipated in the future.  
Direct, nonperturbative
information on the wave functions may also be found 
using instanton models of the QCD vacuum
\ci{Sh-93}.

The traditional approach to derive extra, nonperturbative
knowledge on wave functions 
has been the use of QCD sum rules \ci{Sh-79} to determine their moments of
with respect to $x$.  We shall discuss light-cone wave functions only,
but we note that sum rules have recently been applied to wave
functions with transverse degrees of freedom \ci{Le-94,Zh-94}.

{\bf 2.3.2 Sum Rules for Wave Functions.}
QCD sum rules \ci{Sh-79} have many applications, whenever a nonperturbative
quantity can be related by
analyticity to the integral of a Green function (vacuum expectation
value of a time-ordered product of local fields) over a range of highly
virtual momenta. When this is the case, perturbation theory,
supplemented by the operator product expansion (OPE), may be used to
calculate the integral of the Green function, from which the value of the
matrix element may then be inferred.  

\begin{figure}[htbp]
\includegraphics[angle=90,width=4in]{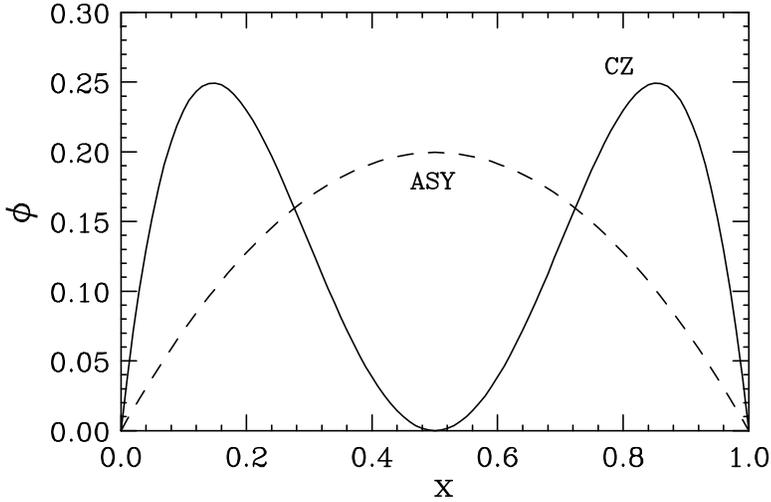}
\caption{CZ wave function (solid) and asymptotic wave function (dashed).
}\label{wffig}
\end{figure}

In the following, we show how QCD sum rules 
may be used to obtain the moments of wave
functions, parameterized in terms of experimentally fitted
gluon and quark vacuum condensates \ci{Sh-79}.  Using this technique, Ref.\ \ci{Ch-84}
obtained the following simple wave function,
\begin{equation}
\phi_{\rm CZ}\left ( x, \mu_0^2\right )
= 5\sqrt{3}f_\pi\; x(1-x)(1-2x)^2\, ,
\label{phicz}
\end{equation}
with $\mu_0\sim 0.5$ GeV.
This result became a common test case for many
subsequent authors. 
This wave function is plotted, along with the asymptotic wave
function, in Fig.\ \ref{wffig}.
Compared to the asymptotic expression 
of Eq.\ (\ref{asywf}), which is centered near $x=1/2$,
the ``CZ wave function" has
a ``double humped" form, with maxima near the extremes of $\xi$.
It has been
the subject of much controversy, as will be discussed in Sec.\  \ref{PFF}.

To derive sum rules for moments of the wave function $\phi_\pi$ 
\ci{Ch-84}, we
first perform a formal Taylor expansion of the quark field $u(z)$ in Eq.\
(\ref{piPsi}),
\begin{equation}
u(z^-)=\sum_{n=0}^\infty {(z^-)^n\over n!}\; (\partial^+)^n\; u(0)\, .
\label{expu}
\end{equation}
Substituting the resulting expression into Eq. (\ref{phidef}), and
carrying out the $z^-$ integrals, we derive the following expansion in
local operators,
\begin{eqnarray}
\phi_\pi(x,\mu^2)
&=& \sum_{n=0}^\infty {(-i)^n\over n!\; {p^+}^{n+1}}
{d^n \delta(x) \over dx^n}\; 
\langle 0|{\bar d}(0) \gamma^+ \gamma_5\; (\partial^+)^nu(0)|\pi(p)\rangle\, .\cr
&\ & 
\label{exp2}
\end{eqnarray}
Moments of $\phi_\pi$ with
respect to
$x$ then pick out individual matrix elements: \cite{Br-80} \cite{Ef-80}
\begin{equation}
(p^+)^{n+1}\int_0^1 dx x^n\phi_\pi(x,\mu^2)
=i^n\langle0|J_n(0)|\pi(p)\rangle\, ,
\label{momentdef}
\end{equation}
where
\begin{equation}
J_n(0)={\bar d}(0)
\gamma^+\gamma_5 (\partial^+)^n u(0)\, .
\label{Jndef}
\end{equation}
Analogous relations between moments of a light-cone distribution and
matrix elements of local operators are familiar from DIS.

\begin{figure}[htbp]
\includegraphics[angle=0,width=4in]{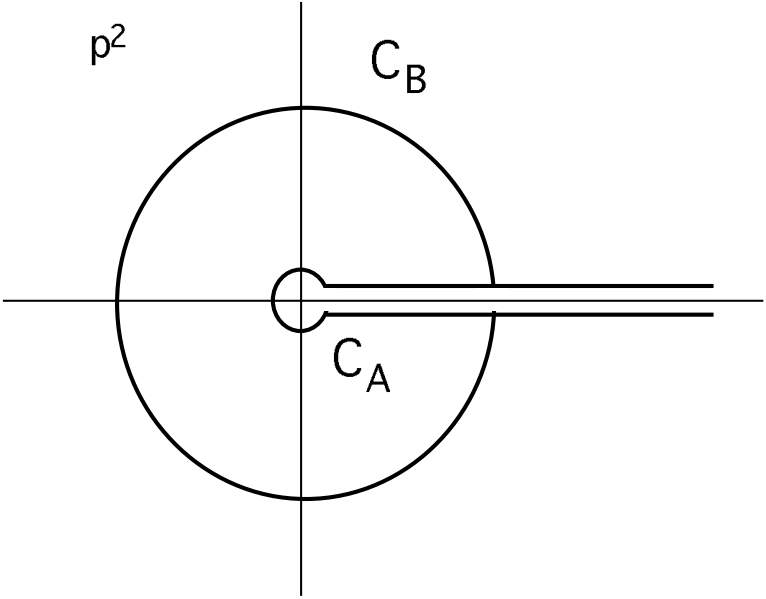}
\caption{Contours in the $p^2$ plane.}
\label{contours}
\end{figure}

We now consider the specific Green function (correlator) 
\begin{equation}
G_n(p^2,p^+)
=
\int d^4z\, {\rm e}^{ip\cdot z}
\langle 0|T\left [ J_n(z)J_0(0)\right]|0\rangle\, ,
\end{equation}
with $J_n (x)$ given by Eq.\ (\ref{Jndef}). Such a two-field Green function
enjoys the analyticity structure shown in Fig. \ref{contours}.  At fixed $p^+ > 0,
G_n (p^2,p^+)$ is an analytic function in the complex $p^2$ plane,
except for poles and branch cuts along the real, positive $p^2 $ axis.  
By
Cauchy's theorem, the integral of $G_n(p^2,p^+)$ along the two contours
$C_A$ and $C_B$ in Fig.\ \ref{contours} give the same result.  The value $p^2 = s_0$
where these contours meet is sometimes called the ``duality interval",
referring to the complementary (dual) manners in which they are
evaluated.

Contour $C_A$ is evaluated using our knowledge of hadron
spectroscopy.  Because the contour runs around the real axis, the
integral is given by the imaginary part of $G_n (p^2,p^+)$ - a sum of delta
function contributions from hadronic bound states
with the quantum numbers of the pion, plus possible
multiparticle continuum contributions.  To emphasize the lowest-lying
state, in this case the pion, we multiply $G_n (p^2 ,p^+)$ by the
``entire" function $e^{-p^2/M^2}$, with $M^2$ an adjustable mass.
Then, a short calculation for the integral along $C_A$ gives
\begin{eqnarray}
I_A(M^2)
&=& \int_{C_A} {dp^2\over  2\pi iM^2}\; G_n(p^2,p^+)e^{-p^2/M^2}\cr
&=&
={\left( ip^+\right)^{n+1}\over M^2}\; \langle x^n\rangle\langle x^0\rangle
+R_A\, ,
\label{isuba}
\end{eqnarray}
where $\langle x^n\rangle$ is the weighted integral of $\phi_\pi$ in Eq.\
(\ref{momentdef}), and
$R_A$ is a remainder, associated with higher-mass resonances such as
the $A_1$, and the continuum.  This integral is referred to 
as a Borel transform.

The corresponding integral $I_B (M^2)$ taken along contour $C_B$ is
computed quite differently.  Along $C_B$, the integrand is evaluated far
from any resonances, and we may hope that it behaves as it does for
Euclidean $p^2 < 0$, where the OPE applies.  Rules for
its calculation are straightforward but rather technical.  The basic
structure of the answer, however, is readily expressed as a sum of a
perturbative term plus two nonperturbative contributions, from
the gluon condensate
$\langle G_{\mu\nu}(A)G^{\mu\nu}(A)\rangle_0$ and 
the quark condensate $\langle {\bar q}q\rangle_0$, 
\begin{equation}
I_B(M^2)
= h_I^{(n)}(M^2,p^+)
+
h_{G^2}^{(n)}\langle G^2\rangle_0 
+h_{({\bar q}q)^2}^{(n)} (\langle {\bar q}q\rangle_0)^2
+R_B,
\label{isubb}
\end{equation}
where $R_B$ represents corrections.  All the $h$'s (``coefficient function
of the OPE)
are computed in perturbation theory.

The values of $M^2$ and $s_0$ are to be chosen to minimize $R_A$
in Eq. (\ref{isuba}).  Values of $<\bar{q} q	>_0$ and $<G^2>_0$ may be found from
the analysis of $e^+ e^- \to$ hadrons \ci{Sh-79}. Finally the coefficient functions
$h^{(n)}$ depend on a renormalization scale $\mu_0{}^2$.  Combining these
choices and parameters, and setting $I_A = I_B$, we may therefore
determine $\langle x^n\rangle$, or equivalently $\langle \xi^n\rangle$, with $\xi = 1-2 x$ the
{\it relative} fractional momentum.  
The CZ wave function in Eq.\ (\ref{phicz}) above was found by fitting
its moments to those found by the sum rules.  

\subsection{Beyond the Pion}
\label{BtP}

{\bf 2.4.1 Generalizations} Most of the developments outlined above for the pion apply as well to
electromagnetic form factors for other hadrons, 
especially baryons \cite{Le-79,Az-80,Du-80,Li-93} and also resonance production,
as well as vector mesons 
and kaons \ci{Le-80,Ch-84,Ji-90,Ag-95}.  The
form factors of baryons are determined by
three-quark valence wave functions, and for both vector mesons
and baryons nontrivial spin structure must be taken into account.   So
long as transverse degrees of freedom may be neglected, however,
spin may be described in terms of conserved helicity, 
where the helicity of a hadron is given
by the sum of the helicities of the its partons.   
A PQCD treatment of violations of helicity conservation
has been proposed in \ci{JR-92,GPR-96}.
We shall
have occasion below to review some of the successes and limitations
of this rich constellation of predictions for hadronic form factors.

For example, the wave function of a proton is a sum of terms 
describing 
total helicity $\pm 1/2$, times functions $\phi_i(x_1,x_2,x_3,\mu^2)$
with $\sum_i x_i=1$.
The application of evolution analysis to these wave functions shows
that asymptotically that have the simple form, 
\begin{equation}
\phi^P_{ASY}(x_i)={\rm const.}\times\; x_1x_2x_3\, .
\label{phipasy}
\end{equation}  
In this case, no readily observed decay amplitude
is available to normalize the asymptotic wave function, and hence
the proton's form factor.
A Sudakov analysis of the proton wave function 
and form factor is also possible,
with the same general properties as for the pion.
\ci{Bo-89,Li-93} It involves
two transverse separations, however, and is correspondingly more
complex.

Another important difference between the proton and the pion is in
the baryonic analogue of Eq.\ (\ref{pifffact}) for helicity form factors $G$ (see below),
which we may represent
schematically as 
\begin{equation}
G(Q^2)
=
\int dx_1dx_2\; dy_1dy_2\; \phi^P(y_j,\mu^2)\; T_G(y_j,x_i,Q)\; \phi^P(x_i,\mu^2)\, ,
\label{protonff}
\end{equation}
where $T_G$, and hence $G$, is proportional to $Q^{-4}$,
and begins at order $\alpha_s^2$.
Here, in contrast to Eq.\ (\ref{pifffact})
for the pion, however, the perturbative expansion of the ``hard-scattering" function $T_G$
receives {\em infrared divergent} contributions from
regions that resemble the Feynman mechanism, in which one quark
carries essentially all of the proton's momentum, beginning at
two loop corrections
\ci{Du-80,Mu-81}.  Such regions are
suppressed by Sudakov corrections.
Progress has been made in quantifying 
this observation for valence PQCD, by introducing transverse 
degrees of freedom for baryons, as for the pion, 
but a complete formalism for baryon form factors, even to leading power
in $Q^2$, remains for the future.

{\bf 2.4.2 Baryon Helicity Matrix Elements.}
For use below, let us define 
electromagnetic  helicity  matrix elements for nucleons.
Taking into account resonance production,
an initial state with helicity $\lambda$ = 1/2
may become a final state with $\lambda^\prime$ = 1/2 or 3/2.
Transitions between a nucleon state ${|N>}$, and final
state ${|N^\prime>}$
can be expressed in terms of  dimensionless helicity matrix elements,
\vs
\begin{equation}
{G_H\ \equiv\ {{1}\over{2M_N}}
<N^\prime, \lambda^\prime|\epsilon^\mu\cdot J_\mu |N, 1/2>}.
\label{GHdef}
\end{equation}
This notation follows \ci{Ca-86}.
The polarization vectors $\epsilon^{\pm,0}$ correspond to right
 ($G_+$) and
left ($G_-$) circularly polarized photons and longitudinally
($G_0$) polarized photons, respectively.
$G_+, G_0, G_-$ describe transitions in which 
$\Delta \lambda$ = 0, 1 and 2, respectively.  
Assuming that helicity is conserved, valence
PQCD suggests that $G_+ \propto Q G_0 \propto Q^2 G_-$. For elastic
scattering, since the recoil nucleon has spin 1/2, only the
helicity conserving  $G_+$ ($\Delta \lambda$ = 0 ) 
and non-conserving $G_0$ ($\Delta \lambda$ = 0) contribute. 

\subsection{Nonasymptotic Form Factors}
\label{NasyFF}

The valence PQCD results above determine the form factor at very high \Q.  How
high one must be, however, is a matter of debate (see below).  
It is therefore important to develop treatments of the transition
to asymptotic behavior.  The evolution of wave functions 
is a step in this direction, but
at moderate \Q, it is necessary to apply methods, or develop models
that take into account processes that are suppressed even by powers 
of \Q\ at high energy.  These include the soft processes
discussed above.  

{\bf 2.5.1 Sum rules for Form Factors.}
Refs.\ \ci{Io-82} and \ci{Ne-82}, have utilized the sum rule approach to directly 
obtain form factors, without the intermediate step
of determining wave functions.  This approach, as described above,
depends on the analyticity properties of Green functions 
that are associated with form factors.  It is thus not
a dynamical theory of soft or hard interactions,
but relies on general properties of QCD, such as the
OPE, in addition to perturbative
calculations.  For a hybrid approach, with features of both
QCD sum rules and valence PQCD, see \ci{Br-94}.

For the pion form factor the relevant Green function may be expressed as
\begin{equation}
T_\mu(p_1,p_2) = i^2\int e^{-ip_1\cdot y+ip_2\cdot z}
\langle 0|T [J (y) J^{em}_\mu(0) J(z)|0\rangle   d^4zd^4y\, ,
\end{equation}
with $J=J_0$ defined as in Eq.\ (\ref{Jndef}), and $J^{em}$ the
electromagnetic current.
In terms of a related scalar amplitude $T$,
this Green function possesses a double dispersion relation,
\begin{equation}
T(p^2_1,p^2_2,Q^2) = {{1}\over{\pi^2}} \int_0^\infty ds_1\int_0^\infty ds_2\, 
{{\rho(s_1,s_2,Q^2)}\over {(s_1-p_1^2)(s_2-p_2^2)}}\, .
\label{ddr}
\end{equation}
The spectral function contains a pion pole which defines the pion
form factor,  $\rho_{\pi\pi}(s_1,s_2,Q^2) = 
2\pi^2f_\pi^2F_\pi(Q^2)\delta(s_1-m_\pi^2)\delta(s_2-m_\pi^2)$,
as well as a continuum above the 3-pion threshold, which also
includes the broad $A_1$ state. 
The form factor  $F_\pi$ is extracted  by 
relating the two contours of Fig.\ \ref{contours},
this time in both
variables $s_1$ and $s_2$.  In \ci{Ne-82}, the Borel transform is
replaced by a simple integral ($M^2\rightarrow \infty$), and $s_0\sim 8\pi^2 f_\pi^2\sim 0.7 $GeV$^2$
is adjusted to reflect this choice, known as ``local duality".  This
leads to a relation between $F_\pi(Q^2)$ and the lowest-order perturbative
contribution to $\rho$, which may be evaluated to give
\begin{equation}
F_\pi(Q^2)
=  1-{\left(1+6s_0/Q^2\right) \over \left(1+4s_0/Q^2\right)^{3/2} }.
\label{NRFpi}
\end{equation}
Expanding in inverse powers of $Q$, this expression behaves as
$Q^{-4}$ for large momentum transfers, and is thus eventually
nonleading compared to the perturbative prediction (\ref{fpiasym}).
Nevertheless, as we shall see, it gives a viable fit to the available
data, which implies at the least that ``soft physics" 
plays an important role in the charged pion form factor at present  
energies.  Beyond lowest order, $\rho$ includes gluonic corrections,
which appear to correspond to the hard gluons of valence QCD.
Similar methods have be used to treat baryon form factors \ci{Ne-83,Ba-96}. 

{\bf 2.5.2 Models.}
Unfortunately the 
complexity of soft processes in QCD
does not lend them to simple physical models. 
Their description
in terms of fundamental QCD is one of the outstanding
theoretical challenges in the theory.
There have, however, been useful attempts to bridge
the low and high \Q\ regions with various phenomenological or
empirical approaches, concentrating on nucleon form factors. 

The  {\em generalized vector dominance model  (VDM)} or {\em hybrid} 
model of \cite{Ga-86a} begins with
the VDM, which yields the requisite low \Q\ form-factor.
Additional
terms join VDM form-factors 
smoothly to PQCD expectations at high \Q\ ($G_M \propto Q^{-4}$
and $G_E \propto Q^{-6}$).
With the appropriate choice of parameters an excellent agreement with
the $G^P_M$ data is achieved over the entire range of available \Q.
Agreement with the other elastic form factors, however, turns out to be
poor in the light of more recent data.

The {\em Constituent quark model} has been   been modified,
and relativized to extend their validity into the few \G\ region of \Q\
\ci{Dz-88,Ch-91,Is-89}.  For example, in the  calculation of  hadronic
form factors in \ci{Is-89}, the constituent quarks, of mass  $\sim$ .33 GeV, have
wave functions which are solutions to  a potential derived
from a quark-quark interaction model. In a light cone frame the wave function
 takes
the form  $\psi(x,{p_T}) \sim X(x)P(x,{p_T})$. The range of $p_T$
in the model wave function effectively has an  ultraviolet cutoff so
that the one-gluon perturbative parts are not included in the derived form
factor. With reasonable choice of $X$
the {\em soft}
components play an important, and even dominant role  over the entire range
of measured \Q. However, there models are not rigorous enough to make
precise predictions.

The {\em diquark model} \ci{An-87,Dz-90,Kr-92}
assumes that
the baryon distribution function can be expressed in terms of two
constituents, a quark and a diquark, which consists of a correlated
quark pair. The diquark structure allows for helicity non-conservation, and thus
at some level can also account for soft processes.
  The diquark becomes completely equivalent to the valence PQCD model
in the high \Q\ limit.
Its several parameters can be tuned to give a good fit 
over the entire range of $G^P_M$, including the transition \Q\ range.

\section{Experimental Status of Hadronic Form Factors}
\label{ESHFF}

\subsection{Pion Form Factors}
\label{PFF}

In this section  we will discuss the $\pi^+$ and $\pi^0$ form factors
as obtained in the reactions $p(e,e^\prime \pi^+)n$   and 
$e^+ + e^- \rightarrow \pi^0$, respectively.  
  Given the relative simplicty of the mesonic
valence state, we might expect perturbative analysis to apply at lower
momentum transfers for pions than for nucleons.  We 
discuss the successes and shortcomings of the valence QCD approach in
explaining the data, and  also point out 
important uncertainties in the
data itself at high \Q.

\begin{figure}[htbp]
\includegraphics[angle=0,width=4in]{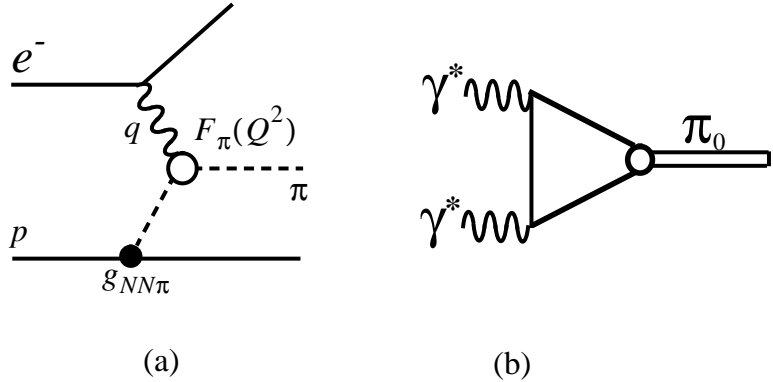}
\caption{ a) Interpretation
of t-channel $\pi^+$ production in terms of the pion form factor $F_\pi$.
b) Lowest order diagram for the $\pi^0$ form factor.}
\label{piplus}
\end{figure}

{\bf 3.1.1 The Charged Pion Form Factor.}
The $\pi^+$ form-factor is obtained by studying 
electroproduction on a hydrogen target
(see Fig.\ \ref{piplus}).  The  aim is to 
separate the ``$t$-channel process", in which the electron scatters from
a nearly on-shell virtual pion emitted from the proton.
This $t$-channel  cross section, which is due to the
exchange of a longitudinal ({\it L}) photon, determines the
pion form factor, though the relation
\begin{equation}
\sigma_L \sim - {{t g^2_{\pi NN}(t)}\over{(t-m_\pi^2)^2}} F^2_\pi(Q^2)\, ,
\end{equation}
where $t$ is the squared momentum transfer to the nucleon, and 
$g^2_{\pi NN}(t)$ is the $\pi NN$ coupling.

\begin{figure}[htbp]
\includegraphics[angle=90,width=4in]{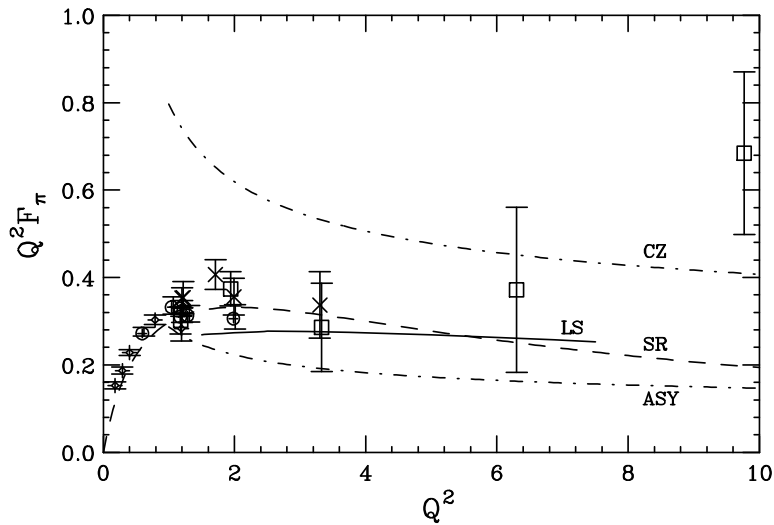}
\caption {The form factor of the $\pi^+$ meson $Q^2F_\pi(Q^2)$ vs \Q.
The data are from {\protect \ci{Be-76a,Be-76b,Be-78}}. See the text for
comments on the interpretation of the higher \Q\ data.
The dot-dash curves labelled CZ and ASY  are obtained from Eq.\ (39),
using $\phi_{CZ}$  and $\phi_{ASY}$ respectively.  The dashed curve labelled
SR is the direct sum rule 
result of {\protect \ci{Ne-83}}. 
 The  solid curve LS is from {\protect \ci{Li-92}}, using the
CZ valence quark distribution amplitude, and including
the effects of Sudakov suppression.}
\label{piplusdata}
\end{figure}

Nearly all the existing high  \Q\ data, shown in Fig.\ \ref{piplusdata},
 were obtained at Cornell
\ci{Be-76a,Be-76b,Be-78}. Care, however, must be
exercized in the interpretation of the higher \Q\ points,
which do not include systematic errors.
The reason for this uncertainty is that the separation of $\sigma_L$
from the complete cross section requires measurements at 
different electron scattering  angles at the same \Q.  This ``Rosenbluth separation'',
  was not practical at the highest $Q^2$ in
this experiment.  For \Q $>$ 4 \G\,  the 
(unwanted) transverse cross section was estimated from an extrapolation
of low \Q\ data, and subtracted by hand.
Thus, although reliable data exist for 
$Q^2 < 3$ \G, the 6.3 and 9.7 
\G\ points provide little help in distinguishing between theoretical models.

There are  also important theoretical issues in the 
extraction of the data. 
For instance, the struck pion is off-shell, and one  must extrapolate
to the physical pion pole at $t = - m_\pi^2$. 
Uncertainties in the $t$ dependence of $g^2_{\pi NN}(t)$ also lead to
uncertainties in $F_\pi$.  
In addition, the reliability of 
high \Q\ form factors extracted in this manner has been questioned
by   \ci{CM-90}, who claim that other {\em hard}, non-resonant
processes compete with the $t$-channel process, and 
may be difficult to separate from it.  
These objections aside, an important future
goal
is to extend the pion form
factor data to higher \Q\ \ci{Ma-94}.

{\bf 3.1.2  Comparison with Theory.}
In the valence PQCD framework, the pion form factor may be
written in factorized form as in  Eq. (\ref{pifffact}).
Treating the hard-scattering at lowest order, with $f_\pi\sim$ 93 MeV, we have
\begin{equation}
F_\pi(Q^2) =  ={16\pi C_F \alpha_s(\kappa^2)\over Q^2}\left|I\right|^2 
\hspace{.25in}{\rm with}\hspace{.25in} I = \int_0^1 dx\, {{\phi_\pi(x)}\over{x}}\, .
\label{fpiI2}
\end{equation} 
This formula, with a valence quark distribution amplitude
derived from QCD sum rules \ci{Ch-84}, denoted   $\phi_{CZ}$,
gives a pion  form factor in rough agreement with the data as shown
in Fig. \ref{piplusdata}.  In obtaining this, 
the variation in $\alpha_S$ was fit to the  evaluated data \ci{PDG-96},
with $\kappa^2$ = \Q/4. 
The asymptotic distribution amplitude $\phi_{ASY}$,
Eq.\ (\ref{asywf}), which yields Eq.\ (20), 
seriously underestimates the data.
  Refering to Fig.\ \ref{wffig}, the difference
 is that $\phi_{CZ}$, Eq.\ (\ref{phicz}), has a ``double-hump" structure,
concentrated near $x\sim 0$ and $1$, and hence yields a larger value for $I$
than the more central $\phi_{ASY}$.
This apparent success inspired many theoretical papers based upon
the ``QCD sum rule" technique for describing exclusive reactions.
The authors of \ci{Is-89,Ra-84} on the other hand, observed that with $\phi_{CZ}$,
Eq.\ (\ref{fpiI2})  is dominated by soft 
 gluon momenta $k^2_g$ (= $xyQ^2$), near  the end-point regions
discussed above.
They  argued that  
Eq.\ (\ref{fpiI2}), or for that matter PQCD,
is invalid in the kinematic regime where data is available,
 because higher-order perturbative corrections would
be uncontrollably large for gluons of such low momenta. If one cuts off
the integral in Eq.\ (\ref{fpiI2}) below a minimum gluon invariant mass,
say $k^2_g \sim .5$ \G, one derives a much smaller 
``legal'' part of the form factor ( $\sim$ 10 - 20 percent remains for \Q\
between 5 and 10 \G\ ).

Roughly, proponents of valence PQCD were faced with the dual problems
of how to keep the main contributions to the $x$ integral in Eq.\ (\ref{fpiI2})
away from the endpoints, at the same time 
enhancing their values relative to the simple use of $\phi_{ASY}$.  
One way of doing this is to resum a selection of higher-order
corrections into the argument of the strong coupling.  Choosing
$\mu^2=xyQ^2$ in $T$ in (\ref{pifffact}) results in a significant
enhancement, because the perturbative running coupling grows as
its scale decreases.  
This running coupling, however,
diverges for $xyQ^2=\Lambda^2_{\rm QCD}$, which requires the
introduction of a scale below which the coupling is ``frozen".
The result is naturally quite sensitive to the cutoff, 
but it can give a reasonable result without dipping too far into
the nonperturbative region \cite{Ji-87}.

In a related development, it was argued that  transverse 
degrees of freedom should not be neglected, and indeed mimic a
gluon effective mass, which suppresses the blowup near $x,y = 0$  \ci{Ha-92}. 
We have already seen how Sudakov resummation of transverse degrees of freedom
in Eq.\ (\ref{listresum}) results in a naturally self-consistent
calculation of the form factor, without cutoffs \ci{Li-92}.
In this case, the CZ distribution amplitudes continued to account
for the existing data, when the enhancement associated
with the running coupling was included. These results for the $\pi^+$ form factor
are plotted in Fig. \ref{piplusdata}. 

The calculation of \ci{Li-92} has been generalized in \ci{Ja-93}, 
who specifically included an ``intrinsic" transverse wave function.  That is, 
in Eq.\ (\ref{listresum})  above, they replaced
  $\exp(-S)\rightarrow \exp(-S)\; \Sigma(x,b)$.
Using a model,
Gaussian shape for $\Sigma$, they found that this further
protected the resulting form factor from the soft region,
but also further supressed the hard part of the form factor
below the data.  Of couse, this 
proceedure introduced an additional parameter, 
in the Gaussian, and it included a constituent
quark mass.

In an alternatative approach (described in Sec.\ \ref{NasyFF}.1 above)
the direct prediction of $F_\pi$ from QCD sum rules,
Eq.\ (\ref{NRFpi}) \ci{Ne-82},  
accounts for most of the measured 
$F_\pi$, without including gluon exchange into its perturbative calculation,
even though the resulting expression decays as $Q^{-4}$ at higher $Q^2$.
In this and other alternatives to the valence quark picture,
the apparent scaling with $Q^2$ of the present data is interpreted as
something of an accident.
The extra contribution of a gluon exchanged between quarks,
which produces $Q^{-2}$ behavior asymptotically, has been estimated 
\ci{Ra-94}, and leads to a modest increase at
the highest available $Q^2$.  These two contributions are
referred to as ``soft" and ``hard" \ci{Ra-94},
the latter being identified with the valence PQCD prediction.

Other publications continue to focus on the relative importance of
soft and hard processes, and in particular how to deal with the
difficult soft sector. Examples are \ci{Br-94,Zh-94,Ki-93,Ja-93,KW-96,Le-94},
 who all conclude that soft processes are 
important for \Q\ corresponding to the existing data.

{\bf Timelike ($s = q^2 > 0$) Pion Form Factor.}
This is obtained  in the reaction 
 $e^+ + e^- \rightarrow \gamma^* \rightarrow \pi^+ + \pi^-$.   
Only one data point exists in the multi-\G\ region, at $s= M^2_{J/\psi}
 \sim 9.6$ \G, obtained from the ratio
 $(J/\psi \rightarrow \pi^+ \pi^- )/(J/\psi \rightarrow e^+ e^-)$ by
 \ci{Mi-93}.
 This point appears to be more reliable than the higher \Q\ space-like
 data.  Its calculation is identical to 
the space-like case in most respects, and
it would be useful to obtain timelike data over a range of \Q.  
  This process was calculated  \ci{Go-95}
with valence PQCD techniques, employing evolution and Sudakov suppression,
as in \ci{Li-92}.  The ratio of experimental
 timelike to spacelike form factors, about 2,
is consistent with the valence PQCD calculation, although 
the overall normalization is low by a factor of two or more,
depending on the light-cone distribution amplitude employed.

{\bf 3.1.3 The $\pi^0$  Form Factor.}
The $\gamma+\gamma^* \rightarrow \pi^0 $ form factor is expected to be a
particularly good test for the 
pion's valence distribution amplitude, since at lowest order in the hard scattering
it is a pure QED processes (see Fig.\ \ref{piplus}b). Higher Fock state
contributions are suppressed by powers of $\alpha_s(Q^2) /Q^2$. 
In addition, there is no analogue of the ``soft", Feynman mechanism contributions, which
require an incoming {\it and} an outgoing pion.

\begin{figure}[htbp]
\includegraphics[angle=90,width=4in]{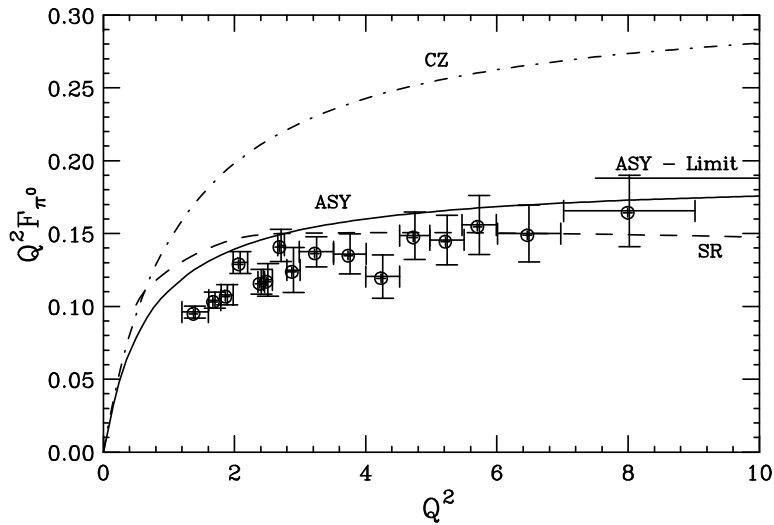}
\caption{Form factor in the few \G\ range for $\gamma+\gamma^* \rightarrow \pi^0$.
Data for \Q\ $>$ 3 \G\ are from
{\protect \ci{Sa-95}}, and lower \Q\ data from {\protect \ci{Be-91}}.  The dot-dash curve
labelled CZ is the result of using $\phi_{CZ}$ in the integral $I$ in
Eq.\ (40), and the solid curve labelled ASY is obtained when using
$\phi_{ASY}$.  To obtain these curves, Eq.\ (40) has been modified
so that the form factor joins smoothly with the known 
value at \Q\ = 0, using a generalization of the prescription of {\protect \ci{Le-80}}.
The horizontal line at the right labelled ASY-limit is the high \Q\
limit, using $\phi_{ASY}$.  The dashed curve labelled SR is the soft form factor obtained
directly from QCD sum rules {\protect \ci{Ra-95}}.}
\label{pizerodata}
\end{figure}

Experimentally, the $\pi^0$  form factor can be studied
via either the Primakoff effect or virtual Compton scattering.
The former is accessible in $e^+-e^-$ colliders, while the 
latter is more appropriate to fixed target machines.
Fig.\ \ref{pizerodata} includes data of the CLEO-II group,
\ci{Sa-95} which reported  measurements up to \Q $\sim$ 8 \G, from reactions
$e^+ e^-\rightarrow \pi+X$. 

Working to lowest (zeroth) order in $\alpha_s$, by analogy with Eq.\ (\ref{fpiI2}),
the relationship between the $\pi^0$ form factor and the valence
quark distribution amplitudes is 
\begin{equation}
F_{\gamma\gamma \pi^0}(Q^2) = {4\over \sqrt{3}Q^2}I\,  , \hspace{1in}
 I = \int_0^1 dx\, {{\phi_\pi(x)}\over {x}} \, .
\label{pi0F}
\end{equation} 
  A calculation \ci{Kr-95} following 
 \ci{Li-92},  including Sudakov effects, and an
intrinsic transverse distribution amplitude as above,
\begin{equation}
F_{\gamma \gamma \pi^0}(Q^2) = \int dx {{d^2b}\over{4\pi}}
 \phi_\pi(x) \Sigma (x,b) \hat{T}_H(x,b,Q){\rm e}^{-S(x,b,Q)}\, ,
\end{equation}
appears to account well for the $\pi^0$ form factor.
The results with $\phi_\pi=\phi_{CZ}$ and  $\phi_{ASY}$ are 
shown in Fig.\ \ref{pizerodata}.  To compare theory with experiment
in this lower \Q\ region, Eq.\ (\ref{pi0F}) has been modified
so that the form factor joins smoothly with the known value at \Q\ = 0,
using a generalization of the prescription of \ci{Br-80}. In this case,
$\phi_{ASY}$ accounts for the data, while  $\Phi_{CZ}$ overshoots it.
A recent  sum rule computation  of  this form factor \ci{Ra-95} also accounts
well for the data.  Because the hard-scattering in the $\pi^0$
form factor starts, as in Eq.\ (\ref{pi0F}), at zeroth
order in $\alpha_s$, the sum rule and valence PQCD approaches 
both begin with the same perturbative diagrams. 
Finally, one may determine the integral $I$ directly by fitting Eq.\ (\ref{pi0F}) to
the $\pi^0$ data, as shown in Fig.\ \ref{pizerodata}. The result 
is close to the value of $I$ for $\phi_{ASY}$.

Comparing the $\pi^0$ and $\pi^+$ form factors, the success of
the asymptotic distribution amplitude in the former suggests that
$\phi_{ASY}$ should be used to compute the 
the valence PQCD contribution to the latter as well.  Then, however,
the valence contribution with lowest order gluon exchange
accounts for less than  one half  of the $\pi^+$ data (see, for instance, \ci{Go-95}).
We conclude that, if the charged pion data is 
accurate at all, either non-valence (soft) contributions, 
or higher-order contributions in valence PQCD,
must play an important role.   We note a lack of need for 
the soft mechanism in the $\pi^0$ form factor, which is consistent
with these observations.  Higher-order hard corrections
are also different in the two form factors, however, so it is difficult
to draw a final conclusion without further study.  
We consider, however, that it is likely that the
soft mechanism plays an important, and possibly dominant,
role in the region of a few \G\ where reliable data exist. 

\subsection{Nucleon Form Factors}
\label{nuclexp}

In this section and the next we consider  nucleon elastic  form factors and
transition form factors involving resonant states of nucleons.
Since there exist stable on-shell baryon targets, the nucleons,
and there are a large variety of final states of spin and isospin, the resonances,
a wealth of experimental information can be accessed. 
Nevertheless,  rather limited data
exist at high \Q.

{\bf 3.2.1  Nucleon Elastic Form Factors.}
The elastic electron-nucleon cross section, expressed in terms
of the {Sachs} form factors, is
\begin{equation}
{{d\sigma}\over{d\Omega_e}}\ =\  
{\sigma_M f_{rec}}\left( {{| G_E |^2\ +\ \tau | G_M |^2}\over {1+\tau}} \ +
\  2\tau | G_M|^2 \tan^2{\theta /2}  \right)   .
\label{Mottsigma}
\end{equation}
Here,
${\sigma_M}$ is the Mott cross section for scattering from a 
point object, ${f_{rec}}$ $= {E^\prime/E}$ is a recoil factor,
$\tau\ \equiv \ Q^2/4M_N^2$ and
$\kappa$ is the anomalous magnetic moment of the nucleon, in nuclear magnetons:
${\rm \kappa_P} = 1.79$, and ${\rm \kappa_N} = -1.91$.

The  helicity matrix elements $G_{\pm,0}$,
defined above in Sec.\ \ref{BtP}, are related to the {Sachs} form factors,
and the {Fermi} ($F_1$) and {Pauli} form factors  ($F_2$)
as follows:
\begin{eqnarray}
 G_+\ &=&\ {{Q}\over{\sqrt{2}M_N}}G_M\ =\ {{Q}\over{\sqrt{2}M_N}}(F_1\ +\ \kappa F_2)\, ,
\cr
G_0\ &=&\ G_E\ =\ F_1\ -\ {{Q^2}\over{4M_N^2}}\kappa F_2\, .
\label{NFFdef}
\end{eqnarray}

For elastic scattering from 
nucleons there are two helicity
conserving and two helicity non-conserving form factors, 
$G^P_M,\, G^N_M$ and $G^P_E,\, G^N_E$, respectively.  At low \Q\ (less than one or two
 \G) all the form factors are consistent with a dipole \Q\ dependence,
$1/(1+Q^2/M^2)^2$, with $M\sim 0.71$ \G . 

At high \Q, valence PQCD predicts that the \Q\ behavior of the helicity
conserving form factors $G^P_M$ or $G^N_M$
should follow $\alpha^2_s(Q^2)/Q^4$
(see Sec.\ \ref{BtP}.1), where we recall that $\alpha_s(Q^2)$ decreases
logarithmically in \Q. 
In addition, their magnitudes are determined by relations like (\ref{protonff})
using nucleon wave functions, 
either of the asymptotic form (Eq.\ (\ref{phipasy})),
or as found, for instance, from sum rules.
 The helicity non-conserving form factors should  $G^{P,N}_E$, should
fall as
$G^{P,N}_M /Q^2$ \ci{Le-80}. 
In the  high \Q\ limit of Eq.\ (\ref{NFFdef}),
 $ F_1 \sim G_M \propto G_+$.
 
 Fig.\ \ref{elasticffdata} summarizes what is known experimentally about these
four form factors, which were mostly obtained at SLAC. $G^P_M$ is known best, followed in
order by $G^N_M$, $G^P_E$ and  $G^N_E$. 
We will consider each  in  turn.
\begin{figure}[htbp]
\includegraphics[angle=90,width=4in]{fig9.epsi}
\caption{ The elastic form factors as a function of \Q\ divided by the dipole
shape: a)$G^P_M$ (spacelike): $\bullet$ - 
{\protect \ci{Wa-94}},  - {\protect \ci{Si-93,An-94}}. b)$G^P_E$.
The curves are:  solid  {\protect \ci{Ji-87}}, dashed  {\protect \ci{Ra-84}},dot-dash  
{\protect \ci{Ga-86a}}.
$G^P_M$ (timelike): $\bullet$ - {\protect \ci{Ar-93}},  - {\protect \ci{Ba-91}}. 
 b)$G^P_E$:  $\bullet$  {\protect \ci{Wa-94}}. The curves are:  dashed  {\protect \ci{Ra-84}}, 
dot-dash  {\protect \ci{Ga-86a}}.
 c)$Q^2F_2/F_1$:  $\bullet$  {\protect \ci{Wa-94}}. 
The curves are:  dashed  {\protect \ci{Ra-84}}, dot-dash - {\protect \ci{Ga-86a}}.
 d)$G^N_M$: $\bullet$  {\protect \ci{Ba-73}}, 
  {\protect \ci{Lu-94}}, $\times$  {\protect \ci{Ro-92}}. The curves are:
  solid  {\protect \ci{Ji-87}}, dashed  {\protect \ci{Ra-84}}, 
dot-dash  {\protect \ci{Ga-86a}}.
 e)$G^N_E$: $\bullet$  {\protect \ci{Ba-73}}, 
 {\protect \ci{Lu-94}}, $\times$  {\protect \ci{Ro-92}}. 
 }
\label{elasticffdata}
\end{figure}

{\bf 3.2.2 The Proton Magnetic Form Factor.} 
Only $G^P_M$ has been measured at high \Q.
In the low \Q\ limit $G^P_E$ and $G^P_ M$
are comparable,  and  can be separated with comparable accuracy by
 a ``Rosenbluth" separation.
Separated form factors 
only exist out to \Q\ $\sim$ 9 \G\ \ci{Si-93}, and unseparated
data exist up to \Q\ $\sim$ 31 \G\ \ci{Ar-86}. However, 
 at lower \Q\ it is observed that
$G^P_E$ is much smaller than  $G^P_M$, and 
that they are roughly proportional. 
Since at higher \Q\ the  $G^P_E$ contribution is  kinematically 
suppressed (see Eq.\ Ref.\ (\ref{Mottsigma})), \ci{Ar-86} estimated $G^P_M$, assuming 
only that $G^P_E$ does not grow anomalously.  The result is
presented as a measurement of  $G^P_M$. 

Much theoretical work has focussed on the application of the valence
PQCD techniques described above to the calculation of the helicity
conserving $G^P_M$ (actually $F_1$) \ci{Ch-84,Ki-87,Ga-87,Ji-87}. 
The broad issues are similar to those
discussed above for the pion.
An advantage relative to the pion, however, is that,
because we can scatter electrons from on-shell protons,
$G^P_M$ is relatively unambiguous  over a larger range of \Q. 
A disadvantage
is that the proton's three valence quarks make it  theoretically
more complex.  

It was observed quite early  that  
the \Q\ dependence of $G^P_M$ is in   
agreement  with quark counting (and hence
valence PQCD) predictions. 
We have already encountered the  basic methods and arguments 
in our  discussion 
for the charged pion form factor.
Once again, calculations based on lowest-order gluon
exchange and asymptotic distribution amplitudes fall far below
the data.  After the initial development of
sum rule distribution amplitudes \ci{Ch-84}, however, the situation
appeared to improve. This was taken as compelling
evidence of the applicability of valence PQCD 
techniques at measurable \Q.
This conclusion has been the focus
of many papers, sometimes quite contentious, by both 
proponents and detractors of
the use of valence PQCD at accessible \Q.

An example is the  calculation of \ci{Ji-87},
based on wave functions
derived from QCD sum rules (Sec.\ \ref{wavefnnp}.3).
The result of using $\phi^P_{CZ}$ from \ci{Ch-84} is plotted along
with the data in Fig.\ \ref{elasticffdata}a.
Once again, the coupling is forced to run with the
virtuality of the exchanged gluons, down to a mass scale
at which it is frozen.  Good fits
to the data were obtained, when this scale (termed an
effective gluon mass) is 0.3 GeV. The curve in  Fig.\ \ref{elasticffdata}a
is not applicable at low \Q, since it is  based on leading order PQCD.

This approach has 
been strongly criticized \ci{Is-89,Ra-84},
for the proton as for the pion.   The basic
question is whether the major contribution to
the form factor comes from gluon exchange at low
virtuality, where higher-order contributions are not under control.
Indeed, sum rules seem to suggest 
asymmetric nucleon distribution amplitudes,
which, as in the case of the pion, enhance contributions from low gluon virtuality.
At the very least, this produces strong sensitivity to
the mass at which the coupling is frozen, and
shakes our confidence in the self-consistency of
the valence calculation.
Also, it was
concluded  \ci{Ec-94,Ec-95} that the uncertainties in obtaining reliable distribution
functions from sum rules, given the experimental uncertainties
in the condensates, are so great
that the distribution amplitudes are essentially undetermined
from sum rules alone.  We may also note that the lattice calculation
of \ci{Ma-89} supports a rather symmetric wave function in the nucleon.

Finally, as for the charged pion form factor, 
the inclusion of transverse momentum effects \ci{Ha-92,Li-92,Li-93}
stabilizes valence PQCD calculations and improves their
self-consistency, while generally reducing them. 
Thus, Ref.\ \ci{Bo-95}, following up on the pion calculation of \ci{Ja-93}, recalculated
$F_1$ (or $G^P_M$) according to the techniques of \ci{Li-92,Li-93}, including
intrinsic $k_\perp$ componants in the distribution amplitude, 
 $\phi_P(x)\rightarrow \phi_P(x)\Omega_P(x,b)$, with $\Omega$ a Gaussian.
In contrast to the pion form factor, the
Sudakov resummation  for the proton form factor 
leaves a sensitivity to large $b$ in a corner of the 
$b,x$ space, necessitating
the inclusion of an ``infrared'' cutoff, that is, a
maximum transverse separation in the distribution amplitude. 
It should be noted, however,
that alternate resummations 
that suppress all large $b$ should be possible, although they have not
been explored in the literature.
As in the pion case, the extended calculation of \ci{Bo-95} reduces the hard scattering
form factor significantly below experimental data, 
for both $\phi^P_{CZ}$ and  $\phi^P_{ASY}$.
 
$G_M^P$ at time-like momentum transfer can extend the range of \Q\ and, together with
space-like data, can further constrain theory.  
The timelike proton form factor has been measured for three values of
 \Q\  near 10 \G\ \ci{Ar-93}.  This, with lower \Q\ data 
\ci{Ba-91} are also shown in Fig.\ \ref{elasticffdata}a.  As in the pion case, 
a factor of about two in the
ratio for the  space-like and time-like form factors is consistent with
expectations from valence PQCD \ci{Go-95}.

{\bf 3.2.3 Proton Electric Form Factor.}    
As \Q\ increases, kinematic suppression of the
contribution of $G_E^P$ in Eq.\ (\ref{Mottsigma}) makes 
a Rosenbluth separation less and less accurate.
As a result, once \Q\ $\sim$ few \G,  the errors on available data for $G^P_E$
are significantly worse than for $G^P_M$.   The most recent data \ci{An-94}
obtained by Rosenbluth separation exhibits much smaller errors than
previous data, and extends the
measured range  out to $Q\sim 9$ \G.
The data, shown  in Fig.\ \ref{elasticffdata}b follow a dipole shape
over the entire range of \Q\,   to within the limited accuracy.

As indicated above, because
 $G^P_E \propto G^P_0$, which is helicity non conserving,
at high \Q\ the ratio $Q^2G^P_E/ G^P_M $  or $Q^2F^P_2/ F^P_1 $ should
approach a constant.  As seen in  Fig.\ \ref{elasticffdata}c, it appears to do so. 
This qualitative success of valence PQCD in the 5 -- 10 \G\ range
makes it attractive to extend the experimental range of accurate  $G^P_E$ data
to higher values of \Q.  A decrease in the ratio
$Q^2G^P_E/ G^P_M $ for large \Q\ of 20 \G, say,
might be a signal that soft processes are still
dominant over hard processes in this range.

Of course, at increasing \Q\ the Rosenbluth separation becomes more difficult.
Other methods, involving polarized beam and target  or
polarized beam and proton recoil polarimeter \ci{Ba-92,Pe-93,Vo-94},
which measure the ratio $G^P_E/ G^P_M$, become 
 more favorable. Using such techniques,
it will be possible to extend measurements of 
$G^P_E$ to higher \Q. 

 {\bf 3.2.4 Neutron Form Factors.} 
Form factors of neutrons are difficult to obtain,
because there are no free neutron targets. Most of the available
data were obtained in quasielastic scattering  from deuterons,
 in which the proton
contribution is subtracted. This method has intrinsic uncertainies, since
one must deconvolute  from the quasielastic peak the contributing neutron
and proton  nuclear 
wave functions, 
which must be independently known,  as well as the  intrusive tails of the
inelastic  processes, which are also broadened by Fermi motion. This becomes 
increasingly difficult  with increasing \Q, as the contribution from 
quasielastic scattering relative to the inelastic processes decreases dramatically. Eventually,  the  tail of
the inelastic background dominates,  and the the
extraction of the quasielastic peak becomes extremely sensitive to 
uncertainties in the modeling of inelastic  processes. Thus, at this time
data on $G^N_M$ and $G^N_E$ are limited to the range \Q\ $\le$ 10 \G\ 
and 4 \G\ respectively.

There are various ways of improving the situation. The
detection in coincidence of the struck neutron along with the electron
can effectively eliminate the quasielastic proton contribution, and significantly
reduce background due to inelastic processes. For $G^N_E$, which is
much smaller than $G^N_M$, polarization asymmetry
techniques can
yield the ratio $G^N_E/G^N_M$.  This method has been employed successfully at lower
\Q\ \ci{Ed-94}, and is currently planned \ci{Ma-93,Da-93} for the few \G\ range.
Neutron form factor data in the \G\ region were obtained at SLAC \ci{Lu-94},
emploing careful Rosenbluth L/T separations of single arm cross section measurements.
The available data are summarized  in  Fig.\ \ref{elasticffdata}c and Fig.\ \ref{elasticffdata}d.  
The  $G^N_M$ data are 
consistent with the dipole shape  over the entire range of \Q, athough there
is significant variance between data sets below about 1 \G.  The data for
$G^N_E$ is consistent with zero up to the highest \Q, although the errors
are quite large.  

The value of this data, even though the range is mostly limited to
the  region where soft processes may still dominate,  is quite
apparent.   All the theoretical non-valence PQCD curves deviate from the data
on $G^N_M$ with increasing \Q. Examples shown are the  hybrid \ci{Ga-86a}
and the QCD sum rule result of \ci{Ra-84}. The constituent quark \ci{Ch-91}, and
vector dominance  \ci{Ho-76} models also appear to diverge monotonically
with increasing \Q.
The data on  $G^N_E$ clearly eliminates the hybrid model, whereas the VDM and
QCD sum rule based calculations are consistant with zero over the \Q\ range. 

To make further use of the selectivity of the nucleon form
factors, it will be important to obtain data on
$G^N_E$, $G^N_M$ and $G^P_E$ at \Q\ greater than the present limits.
Such experiments for   $G^N_M$ and $G^P_E$ have been proposed
for future facilities \ci{BrW-94,Vo-94}.  For $G_M^P$ at least,
its scaling as $Q^{-4}$ over such a large range suggests that
valence PQCD is relevant to its description.   The soft mechanism may,
however, also play an important role, especially at moderate $Q^2$.
The clarification of this role is an important project for theory 
and experiment.

Only  global tests
involving all available form factors can hope to seriously 
select among varying points of view. We stress the importance of measuring
the helicity non-conserving form factors to as high \Q\ as possible,
since in valence PQCD they are driven by non-leading processes, and therefore offer
important constraints on the relative importance of soft and
hard processes with varying \Q.

\subsection{Baryon Resonance Amplitudes and Form Factors}

\begin{figure}[htb]
\includegraphics[angle=-90,width=4in]{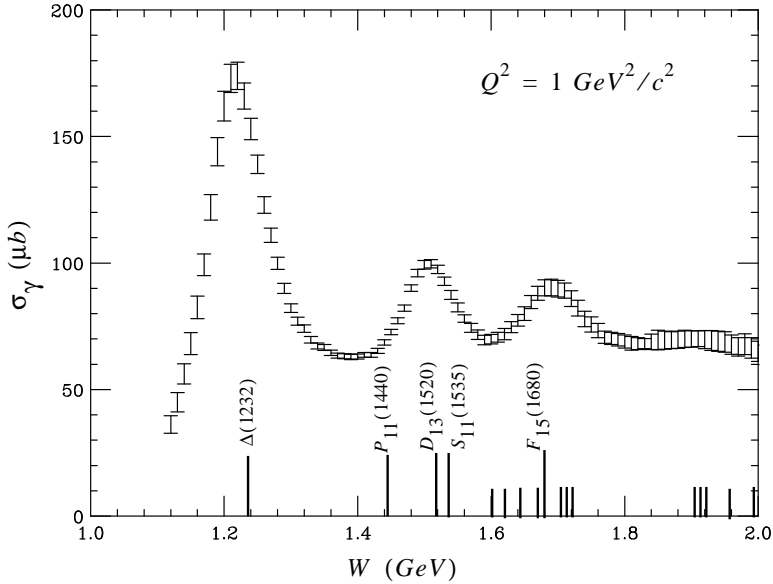}
\caption{ The virtual photon cross section for proton excitation
at \Q = 1 \G. The data are reconstructed from an evaluation by {\protect \ci{Br-76}}.
The known contributing states are indicated at the bottom by vertical
lines. The largest contributing states
are indicated by long vertical lines.}
\label{resonancedata}
\end{figure}

The study of
transition form-factors to excited baryons at high \Q\ can make
an important contribution to  our knowledge of hadronic structure. 
 Fig.\ \ref{resonancedata} shows the virtual photon cross section at \Q = 1 \G\ 
as a function of baryon invariant mass $W$.
For $W\ <\ 2$ GeV, the most significant feature 
is the existence of three maxima, known as
the first, second and third resonance regions.
In this interval there are about 20 known resonances. These
are denoted  $L_{2I,2J}(W)$, where $L$ is 
the angular momentum of the single pion decay, and $I$ and $J$ are
respectively the resonance isospin and spin.
However, except for  the first, which is due to the $\Delta$(1232) the
resonances are largely overlapping, even with a significant non-resonant
underlay. In future programs, the separation of the contributing
 electromagnetic multipoles will
require measurement of exclusive reactions such as
${\rm (e,e^\prime \pi)\ and\ (e,e^\prime \eta)}$ to as high \Q\ as possible,
with polarized beams and targets.

The second resonance region
is dominated by two strong negative parity states,
the  $D_{13}(1520)$ and the $S_{11}(1535)$.
At low \Q\ ( $<$ 1 \G ) the $D_{13}(1520)$ is dominant,
whereas at higher \Q\ ( $>$ 3 \G ) the $S_{11}(1535)$  dominates.
 The Roper resonance, the  $P_{11}(1440)$,
has not yet been definitely observed at \Q\ $>$ 0, but
is of considerable interest since there is  speculation
regarding its character \cite{LiB-92}.
In  the third resonance region,
the largest excitation at low \Q\ is the  $F_{15}(1680)$. The
relative strength of the other states is not well determined, especially
at increasing \Q.
At low \Q\ the excitations indicated in  Fig.\ \ref{resonancedata} have been rather 
successfully described in terms of the constituent quark model.

The current experimental situation is that exclusive 
${\rm (e,e^\prime \pi)\ and\ (e,e^\prime}$ $\eta)$ data exist only up
to \Q\ = 4 \G.
Although there is a total absence of exclusive data above \Q =3 \G,
there are inclusive data in the
resonance region obtained mostly at SLAC (see 
references in \ci{St-93}).  Although the statistical accuracy becomes poor
at high \Q, the three peaks near $W$  = 1232, 1535 and 1680 MeV
 remain prominent, with the $\Delta(1232)$
obviously decreasing with increasing \Q\ relative to the other two.  
After subtraction of phenomenological non-resonant backgrounds the peaks
were fit with resonance functions (\ci{St-93})
to extract transverse form factors,
$|G_T(Q^2)|^2 \equiv (|G_+|^2 +|G_-|^2)/2\tau$, where $\tau \equiv Q^2/4M_n^2$. 

\begin{figure}[htbp]
\includegraphics[angle=90,width=4in]{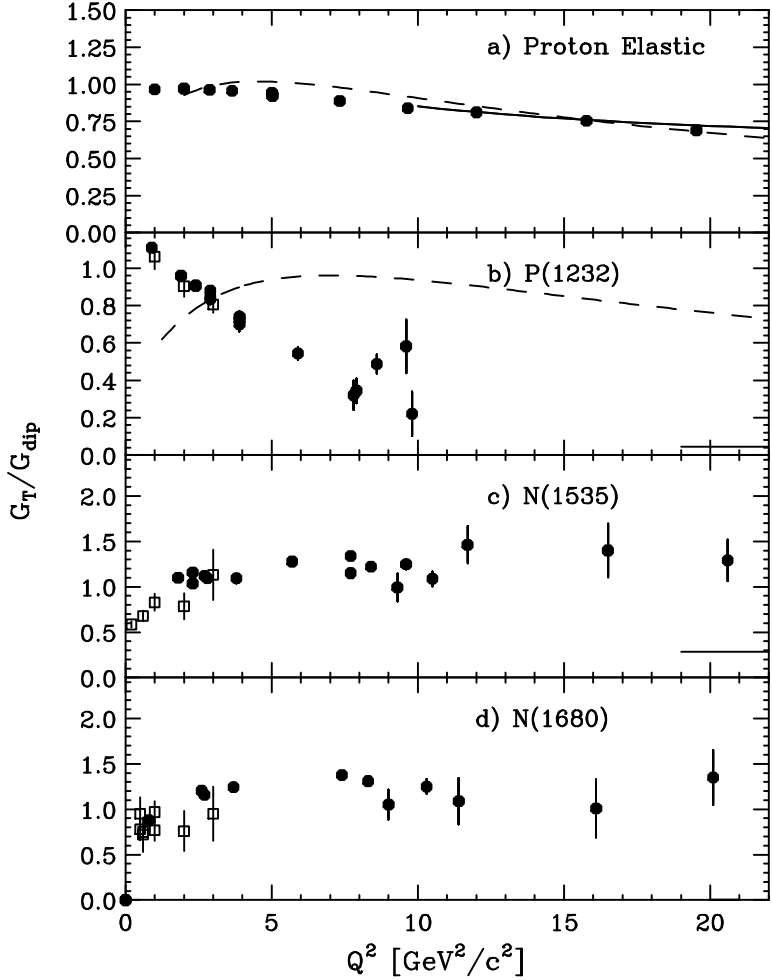}
\caption{The quantity ${G_T/G_{dipole}}$
verses \Q\ for the elastic form factor (a), 
and for transitions to
the first (b), second (c) and third (d)  resonances respectively,
with $G_{dipole} = 2.79(1+Q^2/.71)^{-2}$ in (a)
and $3(1+Q^2/.71)^{-2}$ in (b-d). The first resonance
(b)  is the  $\Delta$(1232)
(the $P_{33}(1232)$). The second resonance ( c) 
at  \Q\ above about 3 \G\ is dominated by the  S$_{12}$(1535).
The third resonance at
low \Q\ is dominated by the F$_{15}$(1680).
The fits for  ${G_T}$ were based on inclusive data referenced in {\protect \ci{St-91,St-93}}.
and selected data from {\protect \ci{Ke-94}}. The elastic proton form factor $G_{Mp}$
is shown in a).
Also shown at lower \Q, denoted by ($\times$), are form factors derived
from amplitudes obtained from exclusive ${(e,e^\prime, p)\pi^0}$
and ${(e,e^\prime, p)\eta}$ data.
The dashed curves are the result of the {local duality} sum rule
calculations of {\protect \ci{Ne-83}} and {\protect \ci{Ba-96}} for the elastic and $\Delta(1232)$
transitions respectively. The solid curve in a) is the  $G_{Mp}$ result
of the PQCD sum rule calcutation of {\protect \ci{Ji-87}} employing $\phi_{CZ}$.
The solid lines at the lower right in b) and c) are the result of the
PQCD calculation of {\protect \ci{Ca-86}} using  $\phi_{CZ}$ for the 
$\Delta(1232)$ and  $S_{12}$(1535)
respectively.}
\label{resonanceffdata}
\end{figure}

The form factors, are shown in  Fig.\ \ref{resonanceffdata},  relative to a dipole
shape. 
Also shown at lower \Q\ are form factors extracted from data
obtained earlier from exclusive ${(e,e^\prime, p)\pi^\circ}$ and
${(e,e^\prime, p)\eta}$ experiments. 

 Fig.\ \ref{resonanceffdata} shows that the form factors obtained for the second and third
resonance regions are consistent with a $Q^{-4}$ dependence, although
with large statistical uncertainty. On the other hand, the $\Delta(1232)$
form factor is decreasing relative to both the elastic as well as 
the second and third resonances. Since this result is obtained from inclusive data 
there are systematic uncertainties in the extraction \ci{Da-95}, and more recent analysis
of the available data \ci{Ke-94} indicates the extent in \Q\ of the decrease is yet resolved.
 
{\bf 3.3.2 $\Delta$(1232): Transition Multipoles }
Since the $\Delta(1232)$ has $J=3/2$ there are three contributing multipoles,
$E_{1+},M_{1+}$, and $S_{1+}$ whose relative contributions are
model dependent. Thus, this is a favorable case for studying models of baryon
structure. 
At low \Q\ in a pure $SU(6)$ non-relativistic CQM  the
$N \rightarrow\Delta $ transition is purely $M_{1+}$ in
character, involving
a single-quark spin-flip with $\Delta L$ = 0. An $E_{1+}$ contribution
is not permitted, since the $\Delta$ and $N$ are both in $L = 0$ states,
which cannot be connected by an 
operator involving $L\ >\ 0$. The addition of a residual quark-quark
color magnetic interaction adds higher $L$ components to the $\Delta$
wave function, and thus introduces
a small $E_{1+}$ component, of perhaps a few percent.
At \Q = 0 the experimental data supports the constituent quark model
prediction
of $M_{1+}$ dominance extremely well. Recent data \ci{Be-97} bears this out.
The data from from Mainz \ci{Be-97} reports
a ratio $E_{1+}/M_{1+} = -.025 \pm .002  \pm .002$.
This ratio remains very small up to $Q^2$ about 1 \G, 
beyond which there is very little data. There exist some earlier data
 at \Q = 3 \G\ \ci{Ha-79}, which has been
evaluated by \ci{Bu-92},  suggesting that $E_{1+}/M_{1+}$ is increasing,
but with large errors, ${\rm Re}(E_{1+}/M_{1+})=0.06\pm 0.02 \pm .03$,  
and we must conclude that the magnitude of $E_{1+}/M_{1+}$ at \Q\ = 3 \G\
remains uncertain. Recently \ci{St-97} exclusive data were obtained at CEBAF
at \Q\ = 3 and 4 \G, for the $\Delta(1232)$ and $S_{11}(1535)$, but at the
time of writing the analysis is not complete. 

At high \Q, valence
PQCD predicts that only helicity concerving amplitudes should contribute. 
The multipole amplitudes for single pion production
may be expressed in terms of helicity conserving
and non-conserving amplitudes
as follows:
\begin{eqnarray}
\Delta\lambda\ &=&\ 0\ :\ A_{1+}\ =\ (3/2)\; M_{1+}\ +\ (1/2)\; E_{1+} \cr
\Delta\lambda\ &=&\ 2\ :\ B_{1+}\ =\ E_{1+}\  -\ M_{1+} \cr
\Delta\lambda\ &=&\ 1\ :\ C_{1+}\ =\ (2Q^2/p^*_\pi)\; S_{1+}\, ,
\end{eqnarray}
where $p^*_\pi$ is the c.m.\ pion momentum.
Thus, helicity conservation implies $B_{1+}\ =\ 0$, or $E_{1+}\ =\ M_{1+}$.
This is quite different from the low \Q\  situation.

QCD sum rule techniques were applied in \ci{Ca-88}, to calculate the 
distribution functions for the $\Delta$(1232) excitation.
The CZ \ci{Ch-84}  proton wave function yields a small transition form factor,   
$Q^4G_T(Q^2)\sim 0.07$ asymptotically. This can be traced to a cancellation 
in the leading order term  of the matrix elements
connecting the symmetric $\Delta$(1232) distribution amplitude, with the
symmetric and antisymmetric
proton distribution amplitude respectively. Schematically, $|\langle\phi_\Delta 
|T_H| \phi^P_S\rangle\ +\ \langle \phi_\Delta |T_H| \phi^P_A\rangle|$ is much smaller than
either alone.

If the leading amplitude of the P $\rightarrow$ $\Delta$ 
transition is
indeed small, the anomalous shape of the  transition form factor might
be explained as follows. 
At high \Q, the leading order helicity conserving amplitude
dominates over
the helicity non-conserving amplitude. That is, $A_{1/2}>> A_{3/2}$.
A suppression of the  $A_{1/2}$ 
amplitude at all \Q\ due to the cancellation of the
symmetric and antisymmetric matrix elements, might then result
in the dominance of the  $A_{3/2}$ 
amplitude over a larger
range of \Q\ than otherwise expected, and ${Q^4G_T(Q^2)}$ would decrease
as a function of \Q .
In fact the evidence that $E_{1+}/M_{1+}$ is still small
for \Q\ up to 3 \G, is consistent with the dominance of non-leading
processes.

Recently \ci{Ba-96} the local duality procedure was applied to the
$\Delta(1232)$ form factor, and it was found, as in the pion case, 
that the form factor in the few \G\ region can be accounted for  
by purely soft processes (see  Fig.\ \ref{resonancedata}). However,
it then falls significantly
below the experimental values at higher \Q, which might be evidence
that hard processes are playing an increasing role. 

It will be interesting in the future to determine whether ${Q^4G_T(Q^2)}$ does
indeed level off above  \Q = 10 \G, and
where the $E_{1+}$ amplitude becomes comparable to the $M_{1+}$. This would
support the valence PQCD description.

{\bf 3.3.3  The Second Resonance.}
 Fig.\ \ref{resonanceffdata} shows that at high \Q\ the form factor for  
the peak at W $\sim$ 1535 MeV approaches the $Q^{-4}$ dependence predicted by valence PQCD. 
Although the  $D_{13}(1520)$ is dominant at \Q\ = 0, there is a
crossover and  the $S_{11}(1535)$ dominates  the $D_{13}(1520)$ at \Q\ $\sim$
few \G\ \ci{Ha-79}. Another unique feature of the  $S_{11}(1535)$
is that it is the only excited state with a large $\eta$ decay branching
ratio ($\sim 50\%$), so that experimentally it is easily is isolated.  

Ref.\ \ci{Ca-88} presents a calculation of the ${ proton \rightarrow\ S_{11} }$ transition
form factor in the valence PQCD framework. 
The result is a behavior similar  to the elastic form factor.  Although the results
are about a factor of two lower than the data, the authors remark that
theoretical uncertainties in the distribution functions, and higher order
contributions to $\alpha_s$ are probably great enough to account for these
discrepancies.

{\bf 3.3.4 The Third Resonance. }
  Fig.\ \ref{resonanceffdata} shows that at high \Q\ the form factor for the peak near W = 1680 MeV
is consistent with the predicted $Q^{-4}$ behavior. The
errors are large, however, and it is not clear how many resonances are
contributing to this peak.
 The potential for obtaining separated resonance amplitudes
at high \Q\ with exclusive reactions is very good. This is particularly true
since it has been demonstrated \ci{Bl-70,St-91,Ca-90}, that the non-resonant
background diminishes with \Q\ at approximately the same rate as the
resonances.

{\bf 3.3.5 Duality.} A very interesting concept is that of duality between
resonances and the non-resonance continuum in the $W$ region where they
overlap. One observes  \ci{Bl-70} that the rate of decrease with \Q\ of
the resonance cross sections approximately follows the extrapolation
of deep-inelastic scaling into the resonance region, suggesting
that both processes are related by the same underlying
physics. Later, this was put on firmer ground, and it was shown
that leading logarithmic corrections  extend the duality range
in \Q\  \ci{Ca-92,Ca-95}. However, all of this is based on 
analyses of inclusive data, which cannot effectively separate
non-resonance from resonance contributions.  In order to access 
this very fundamental result one really needs to 
have a clean separation of resonance and non-resonance data over a
large range of \Q, which can only be accomplished by the measurement
of exclusive reactions.

\section{Conclusions}

As seen, much work remains in both experiment and theory. 
Valence PQCD and factorization appear to
be an  attractive starting point for treating high \Q\
form factors, although how high \Q\ must be for
valence PQCD to dominate
remains controvertial, and most probably 
depends on the specific reaction.  Opinions on this matter vary strongly, from
those who maintain  that the required \Q\ is much higher than is
likely to be experimentally accessible in the forseeable future, to those who
believe that valence PQCD is already applicable at \Q\ as low as a few \G. 
We suggest that the quality and extent of existing data
does not allow a definitive conclusion,  but 
that soft non-perturbative processes probably play an important
 role for much of the existing data. 
On the other hand, given the complexity of QCD, there is a need for further
theoretical work,  based on  fundamental principles of QCD, 
to deal with the soft, or Feynman mechanism. 
Indeed, it may be possible to express form factors in the transition
region as a sum over valence PQCD and Feynman mechanism contributions,
with the hard-scattering imbedded in the latter treated with PQCD
methods.  For the truly soft region,
lattice calculations may play an increasing role in the future \ci{Ma-89,Ne-95}.  

On the experimental side, one must push the frontiers to as high \Q\ as
technically feasible, to provide data which has the best chance of
testing these ideas. One should also go beyond the experiments which merely
test constituent scaling, to those which test other central tenets
of theory, such as helicity conservation. Such work has now begun at
CEBAF, and may be further extended by a proposed European facility, ELFE
\ci{Ar-92}
In summary, this area appears to offer some of the most interesting  theoretical
and experimental challanges for the next decade. 

ACKNOWLEDGEMENTS: The authors wish to acknowledge and thank  P. Bosted
 and A. Radyushkin for their  valuable assistance. This work was supported
in part by the National Science Foundation under grants PHY-9507412 and
PHY-930988.

\end{document}